\documentclass[aps,pra,twocolumn,tightenlines,superscriptaddress,amsmath,amssymb, longbibliography]{revtex4-1}
\usepackage{graphicx}
\usepackage{bm}
\usepackage{accents}
\usepackage{multirow}
\usepackage{grffile}
\usepackage{amsfonts}

\usepackage{xcolor}
\definecolor{dark-red}{rgb}{0.4,0.15,0.15}
\definecolor{dark-blue}{rgb}{0.15,0.15,0.4}
\definecolor{medium-blue}{rgb}{0,0,0.5}
\usepackage{hyperref}
\usepackage[all]{hypcap}
\usepackage{cleveref}
\hypersetup{
    colorlinks, linkcolor={dark-red},
    citecolor={dark-blue}, urlcolor={medium-blue}
} 

 \newcommand{\be}{\begin{equation}}
 \newcommand{\ee}{\end{equation}}
\newcommand{\bea}{\begin{eqnarray}}
\newcommand{\eea}{\end{eqnarray}}
\newcommand{\ba}{\begin{eqnarray*}}
\newcommand{\ea}{\end{eqnarray*}}

\newcommand{\m}[1]{\mathcal{#1}}

\usepackage{multirow}

\newcommand{\ket}[1]{\lvert\, #1\, \rangle}
\newcommand{\bra}[1]{\langle\, #1\, \rvert}
\newcommand{\degen}{\Omega}

\DeclareMathOperator{\tr}{Tr}

\DeclareMathOperator{\arcsinh}{arcsinh}
\newcommand{\erg}{e}

\newcommand{\g}{g}

\newcommand{\ratio}{\varrho}
\newcommand{\frerg}{F}

\newcommand{\scl}{{\ell}}

\newcommand{\Mm}{{M}}
\newcommand{\mm}{{m}}

\newcommand{\action}{{S_\ell}}
\newcommand{\sB}{\mathcal{B}}
\newcommand{\dif}{d}
\newcommand{\ssp}{\hspace{0.4pt}}
\newcommand{\normb}[1]{\big\lvert #1 \big\rvert}
\newcommand{\proj}[1]{\ket{#1}\!\bra{#1}}
\newcommand{\dt}[1]{\accentset{\vspace{0.5pt}\hspace{0.5pt}\mbox{\large .}}{#1}} %

\usepackage{color}
\usepackage[english]{babel}
\usepackage{dcolumn}

\usepackage[caption=false, subrefformat=parens, labelformat=parens]{subfig}

\begin{document}

\title{Scaling analysis and instantons for   thermally-assisted tunneling and Quantum Monte Carlo simulations}

\author{Zhang Jiang}
\affiliation{QuAIL, NASA Ames Research Center, Moffett Field, California 94035, USA}
\affiliation{SGT Inc., 7701 Greenbelt Rd., Suite 400, Greenbelt, Maryland 20770}
\author{Vadim N. Smelyanskiy}
\affiliation{Google, Venice, CA 90291, USA}
\author{Sergei V. Isakov}
\affiliation{Google, 8002 Zurich, Switzerland}
\author{Sergio Boixo}
\affiliation{Google, Venice, CA 90291, USA}
\author{Guglielmo Mazzola}
\affiliation{Theoretische Physik, ETH Zurich, 8093 Zurich, Switzerland}
\author{Matthias Troyer}
\affiliation{Theoretische Physik, ETH Zurich, 8093 Zurich, Switzerland}
\affiliation{Quantum Architectures and Computation Group, Microsoft Research, Redmond, Washington 98052}
\author{Hartmut Neven}
\affiliation{Google, Venice, CA 90291, USA}

             
\begin{abstract}

We develop an instantonic calculus to derive an analytical expression for the thermally-assisted tunneling decay rate of a metastable state in a fully connected quantum spin model. The tunneling decay problem can be mapped onto the Kramers escape problem of a classical random dynamical field. This dynamical field is simulated efficiently by path integral Quantum Monte Carlo (QMC). We show analytically that the exponential scaling with the number of spins   of the thermally-assisted quantum tunneling rate and the escape rate of the QMC  process are identical.  We relate this effect to the existence of a dominant instantonic tunneling path.
The instanton trajectory is described  by nonlinear dynamical mean-field theory equations for a single-site magnetization vector, which we solve exactly.  Finally, we derive scaling  relations for the \lq\lq spiky" barrier shape when the spin tunneling and QMC rates scale polynomially with the number of spins $N$ while a purely classical over-the-barrier activation rate  scales exponentially with $N$.
 \end{abstract}
  \maketitle

\section{Introduction}
   
Computationally hard combinatorial optimization problems can be
mapped to classical spin glass models in statistical
physics~\cite{fu_application_1986}. The energy
landscape of the corresponding spin Hamiltonians $H_P$ possesses a large number of spurious local minima.  
Classical optimization strategies, such as simulated annealing (SA),
exploit thermal over--the--barrier transitions for the search
trajectory through the energy landscape towards low-energy spin
configurations. In quantum optimization algorithms, such as quantum annealing (QA)~\cite{PhysRevE.58.5355,*brooke_quantum_1999,Farhi20042001,Santoro29032002} (see also \cite{boixo2014,*Ronnow:2014fd,*King:2015vl} for recent results) tunneling can play a functional role by providing additional pathways to low-energy
states~\cite{boixo_computational_2014}. 
  
In an archetypical example of QA in spins models, the state
  evolution is determined by a time-dependent Hamiltonian
  $H(t) = H_P - \Gamma(t) \sum_j \sigma_j^x$, where $\sigma_j^x$ is a
  Pauli operator for the $j$ spin, and $\Gamma(t)$ slowly interpolates between
  a large value and $0$. At sufficiently small values of $\Gamma$, all
  low-energy eigenstates of $H(t)$ are localized in the vicinity of
  the minima of $H_P$~\cite{Altshuler13072010} (pure states in the
  spin-glass models). In QA the energy landscape is time dependent and
  the energies of two minima can exchange orders. Due to the tunneling between the minima, this results in an avoided
  crossing with the energy gap $\Delta$.
  
In the absence of an environment and for sufficiently slow evolution, QA corresponds to an adiabatic evolution where the
  system closely follows the instantaneous ground state of
  $H(t)$~\cite{Farhi20042001}. The state dynamics can be described as a cascade of Landau--Zener transitions at the
  avoided crossings.  
 
The interaction with an environment can suppress tunneling between two given states of the system.   The rate of incoherent tunneling decay is $W  \propto  \Delta^2/(\hbar^2\gamma)$ when the relaxation rate $\gamma$ due to environment is much larger than the energy gap, i.e., $\gamma \gg \Delta/\hbar$. However, the environment also gives rise to thermal excitations to higher-energy levels from where the system can tunnel faster~\cite{kechedzhi_open-system_2016}. This is called thermally assisted
 tunneling~\cite{larkin1983pis,*garanin1997thermally,*affleck_quantum-statistical_1981},
 and has recently been discussed in applications to flux qubit
 QA ~\cite{amin_macroscopic_2008,dickson_thermally_2013}.
 
We assume that the system with the microscopic
Hamiltonian $H(t)$ has a free-energy minimum associated with a thermodynamically
metastable state, and that the incoherent tunneling decay rate $W$ of this state is much smaller than the smallest rate of relaxation $\gamma$ towards the quasi-equilibrium distribution in the domain associated with this state.

Consider an eigenstate $|\psi_n \rangle$ of $H(t)$
localized in the metastable domain.  The energy of  $|\psi_n \rangle$ acquires an imaginary part, $E_n-i\hbar W_n/2 $, due to incoherent tunneling. Thus, the partition
function of the metastable state becomes a complex number $Z_0
 = \sum_n e^{- \beta(E_n - i \hbar W_n/2)/\hbar} $, where $\beta/\hbar$ is the inverse
 temperature.  For  $\hbar W_n \ll
 E_n$ we expand the partition function in $W_n$ and express the total tunneling decay rate $W$
of the metastable state in the form
\begin{equation}
  W =-\frac{2}{\beta}\frac{{\rm Im}(Z_0)}{{\rm Re}(Z_0)}= \frac{\sum _n W_n  e^{- \beta E_n/\hbar}}{\sum_n e^{- \beta E_n/\hbar}}\;, \label{eq:F}
 \end{equation}
where $ {\rm Re}(Z_0)$ is computed by neglecting the imaginary parts of the energies. The quantity $W$ introduced above is given by the average over the micro-canonical decay rates $W_n$
weighted with the corresponding (quasi) equilibrium populations.
Due to the entropic effects, the decay rate $W$ can have a very steep and non-monotonic dependence on temperatures \cite{kechedzhi_open-system_2016}. It can exceed the zero temperature rate or even the coherent tunneling frequency. Therefore, optimal
protocols of QA must also explore the finite-temperature
regime.

Tunneling transitions represent the bottlenecks of
QA~\cite{knysh_zero-temperature_2016} since the tunneling rate $W$ decreases
exponentially with the size of the domain $D$ of cotunneling spins
(typically, Hamming distance between the minima),  $W = B_D
e^{- D\, \alpha}$, where $B_D$ is a polynomial prefactor.  

In this paper, we investigate analytically  thermally assisted tunneling as a {\it quantum} computational resource for an arbitrary temperatures. We compare the scaling exponent $\alpha$ in the tunneling rate $W_{\rm tunn}$ to the
corresponding transition rate of path integral Quantum Monte Carlo (QMC), which is a classical algorithm that is often used to simulate QA in
spin glasses~\cite{Santoro29032002,PhysRevB.66.094203,*martonak_quantum_2004,*battaglia_optimization_2005,*santoro_optimization_2006,boixo2014,Heim12032015}, and it is the most efficient algorithm to compute exponentially small gaps at first-order phase transitions~\cite{young2010first,hen2011exponential,boixo2014}. 
We also provide numerical evidence to support the theoretical findings. The numerical results address  the case of non-zero bias and finite (but  small) temperatures,  expanding the numerical findings  presented in Ref.~\cite{isakov_understanding_2016}


We demonstrate in a closed analytical form for mean-field quantum spin models in transverse field 
that  the transition rate $W_{\rm QMC}$ of QMC and the quantum tunneling rate $W_{\rm tunn}$ have identical exponential scaling  with number of co-tunneling spins . 
This finding applies to the situations where the tunneling 
is dominated by a most probable path (an instanton). We will employ
the path-integral formalism, reminiscent of the one used in the case of
continuous~\cite{coleman_fate_1977,caldeira1983quantum,larkin1983pis}
and spin~\cite{owerre2015macroscopic} systems, such as tunneling of magnetization in
nano-magnets~\cite{chudnovsky2005macroscopic}. Remarkably, despite
the big body of literature in this topic, spin tunneling has only
been studied, using path integrals, in systems with fixed total spin. We will
introduce a non-perturbative spin path-integral instanton calculus
for systems where, due to thermal fluctuations, the
total spin is not preserved. We augment our analytical results by detailed numerical studies.

Finally, we  compared the quantum tunneling rate, or equivalently the classical QMC rate, with the transition rate corresponding to a  classical, purely thermal activation over the barrier. We considered the case  where  the barrier has a form of a tall and narrow spike. We assumed that the  width and height of the spike scale at most linearly with the number of spins $N$. We found the relations for the parameter range of the  two  scaling exponents  where the quantum tunneling rate (and the classical QMC rate) do not scale exponentially with $N$ while the rate of over-the-barrier thermal escape does. 

In Sec.~\ref{sec:WKB}, we review the  results for  a  thermally assisted quantum spin tunneling  rate  initially obtained in \cite{kechedzhi_open-system_2016} using the Wentzel-Kramers-Brillouin (WKB) approach. In Sec.~\ref{sec:KramersQMC}, we developed a Kramers theory of the escape  rate for the QMC tunneling simulations. In Sec.~\ref{sec:comp}, we use a path integral approach to establish a detailed connection between the WKB results for the quantum tunneling rate and the associated instanton  trajectory and QMC escape rate. In Sec.~\ref{sec:QMCnum}, we provide the results of numerical studies. In Sec.~\ref{Sec-spike}, we apply the theory developed in this paper to the tunneling problem with small and narrow barriers.

\section{\label{sec:WKB} Thermally assisted tunneling in multispin systems}
\subsection{ Wentzel-Kramers-Brillouin (WKB) approach}

The thermally assisted tunneling rate $W_{\rm tunn}$~\eqref{eq:F} in the mean-field models where the total spin quantum number is not conserved was analyzed in
Ref.~\cite{kechedzhi_open-system_2016} using the discrete
WKB approach~\cite{Garg-wkb}. We shall
summarize below the results in Ref.~\cite{kechedzhi_open-system_2016} for the archetypical model of 
a quantum ferromagnet in an $N$-spin
system~\cite{Semerjian-wkb,kechedzhi_open-system_2016}
\begin{align}
 \hat H = -2 \Gamma \hat S_x - N g(2 \hat S_z/N), \quad \hat S_\alpha
  = \frac 1 2 \sum_{j=1}^{N} \sigma_{\alpha}^{j}  \label{eq:H-sigma}\;,
\end{align}
where $\Gamma$ is the strength of the transverse
field; $\sigma_{\alpha}^{j}$ is the Pauli
matrix of the $j$th spin, where $\alpha=x,y,z$; $\hat S_\alpha$ is the $\alpha$ component of
the total spin operator; and $g$ is an arbitrary function of $2 \hat S_z/N$.

The  mean-field interaction energy density $-g(m)$ in (\ref{eq:H-sigma}) has a local and a global
minimum. This class of models is known to have large free-energy
barriers that lead to exponential (in $N$) slowing down for
QA~\cite{jorg2010energy,Semerjian-wkb,kechedzhi_open-system_2016}. 

\begin{figure*}[t]
\begin{minipage}{.45\textwidth}
  \includegraphics[width=0.9\textwidth]{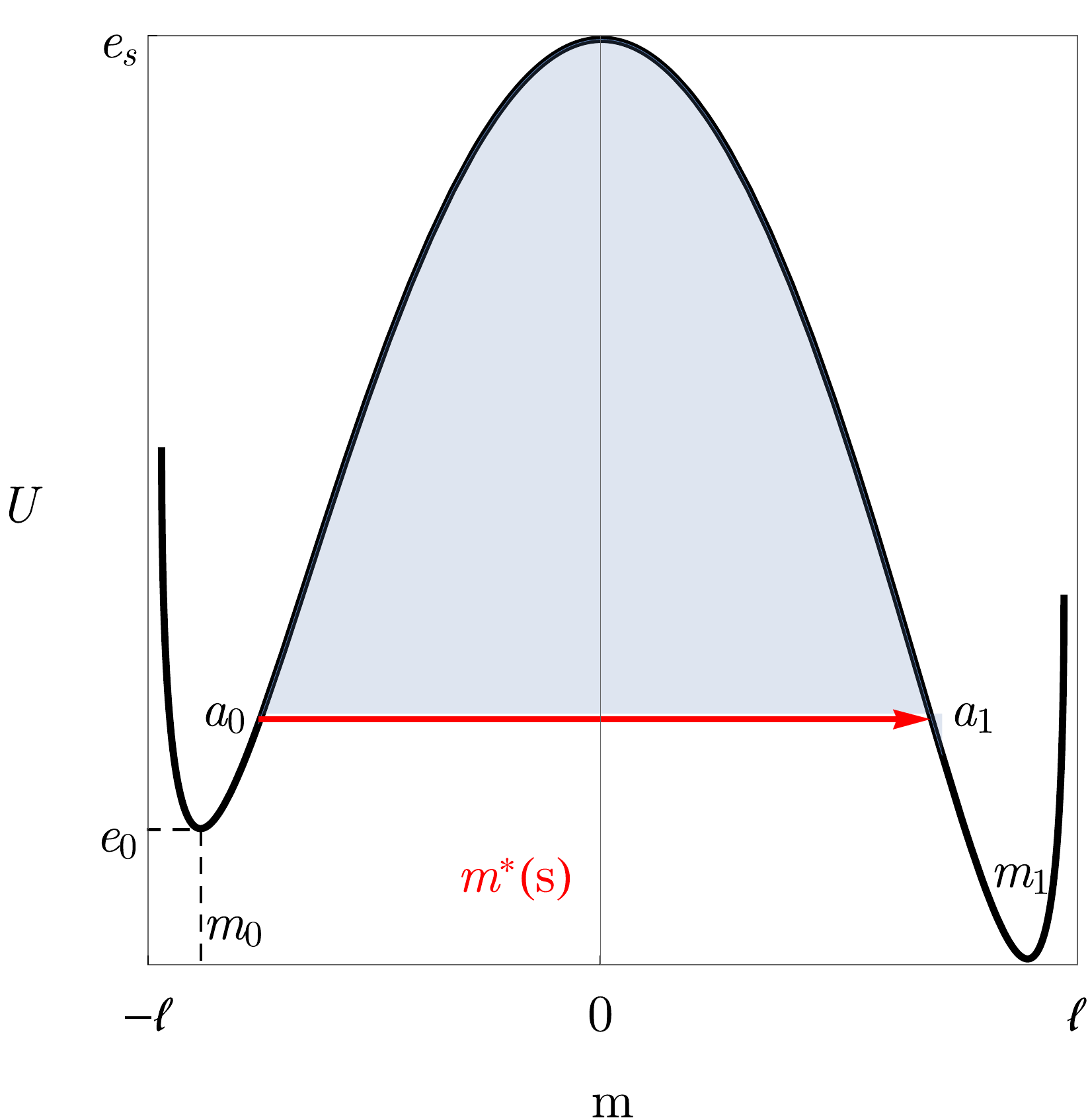}
\end{minipage}
\begin{minipage}{.45\textwidth}
       \includegraphics[width=0.9\textwidth]{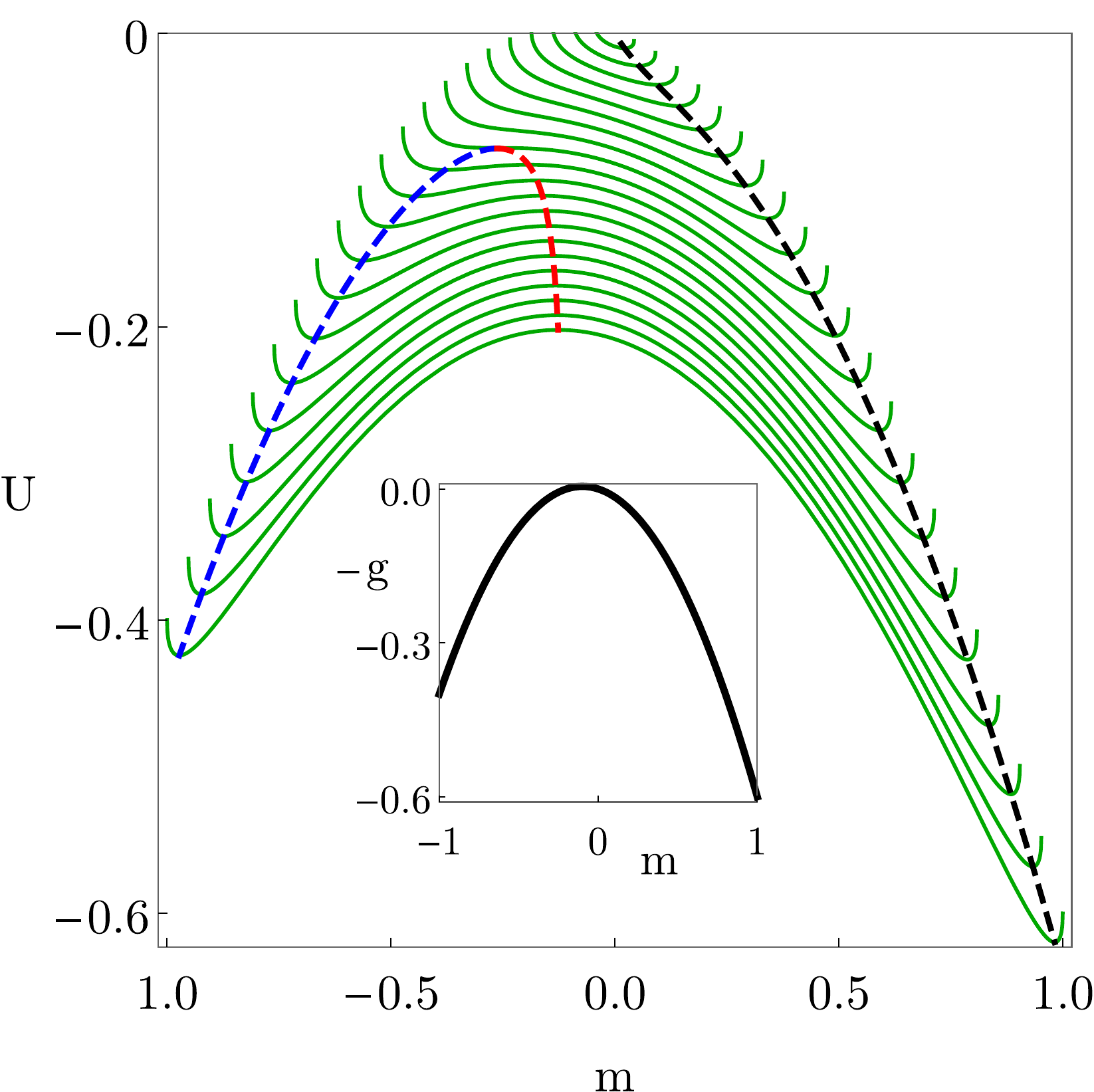}
\end{minipage}
    \caption{(Left)
     Plot of the effective potential
    $U_\ell(m)$ as a function of the magnetization  $m$ for the
      Curie-Weiss model (\ref{eq:CW}) for $\ell=1$, $\Gamma=0.4$, and $h=0.015$. The red
      line depicts the tunneling path (instanton) at zero
      temperature. Barrier energies above the instanton energy are
      shown with blue filling. (Right) Plots of the effective
      potential for the Curie-Weiss model at different values of
      $\ell$ shown with green lines with $h=0.1$ and $\Gamma=0.21$.  The colored dashed  lines
      correspond to the local minimum (blue), the maximum (red), and the global
      minimum (black) of $U_\ell(m)$  for different values of $\ell$. 
      The total spin parameter $\ell$ changes in the range from
      $\ell_c$ to  1 where  $\ell_c=(h^{2/3}+\Gamma^{2/3})^{3/2}$ is
      the smallest value of $\ell$ beyond which the effective
      potential is monostable. The domain of the effective potential
      for each $\ell$ is $m\in(-\ell,\ell)$. The inset in the right figure shows the plot of interaction energy density $-g(m)$. We see by comparison of figures in left and right panels that  for a given $\ell$ the turning point at the barrier exit satisfies the condition $a_1 \leq \ell$ where the equality is reached only for $\ell=1$.
 \label{fig:Veff}}
\end{figure*}

The squared total spin operator $\hat S^2 = \sum_\alpha S_\alpha^2$
commutes with the Hamiltonian $\hat H$~\eqref{eq:H-sigma}. The eigenstates of $\hat H$ can be
expanded in the basis of the operators $\hat S_z$ and $\hat S^2$,
\begin{equation}
\ket{\Psi}=\sum_{S={\rm frac}( N/2)}^{N/2}
\sum_{\Mm=-S}^{S}\sum_{\nu=1}^{ \degen(N, S)} C_{\Mm}^{S,\nu}|\Mm,S,\nu\rangle\label{eq:Psi}\,,
\end{equation}
where ${\rm frac}( N/2)$ is the fractional part of $N/2$, and $\degen(N, S)=\binom{N}{N/2 -S} \,-\, \binom{N}{N/2 -S-1}$ is the number of distinct irreducible subspaces with a given total
spin quantum number $S$.
The coefficients $C_{M}^{S,\nu}$ obey the stationary Schr\"odinger equation
\begin{multline}
-\Gamma \sum_{\alpha=\pm 1} \sqrt{(S+\alpha \,M+1)(S-\alpha\, M)}\,
C_{M+\alpha}^{S,\nu} \\-N g(2M/N) C_{M}^{S,\nu}=E C_{M}^{S,\nu}\,,\label{eq:Sch0}
\end{multline}
where $M = -S,-S+1,\ldots,S$.


In what follows we will study the limit $N \gg 1$ and assume that $S={\cal O}(N)$. We introduce the normalized total spin quantum number $\ell=2S/N \in [0,\,1]$. To exponential in $N$ accuracy, the number of distinct irreducible subspaces  $\Omega_{\ell}$ for a given value of $\ell$ equals
\begin{equation}
\Omega_{\ell}\sim \exp(N Q_\ell),\,\, Q_\ell=\sum_{\alpha=\pm 1}\frac{1+\alpha \ell}{2}\ln\left(\frac{2}{1+\alpha\ell}\right)\,,\label{eq:Q}
\end{equation}
where the binary entropic factor $Q_\ell$ takes the maximum value $Q_{\ell =0}=\ln 2$. To the leading order in $S$, Eq.~\eqref{eq:Sch0} can be written in the form $\varepsilon_\ell(m,\hat p)\,C_{m}^{\ell}=e  C_{m}^{\ell} $ where 
\begin{align}
\varepsilon_\ell(m,\hat p)=-\Gamma \sqrt{\ell^2-m^2}\cos \hat p- g(m) \;,\label{eq:Sch}
\end{align}
 momentum $\hat p=- (2 i /  N) \partial / {\partial m}$, and the rescaled variables are
\begin{align}
  m = \frac{2 M}{N} , \quad \ell = \frac{ 2 S}{N} ,\quad  e =  \frac{E}{N},\quad
  C_{M}^{S,\nu} = C_{m}^{\ell}.\label{eq:rescaled}
\end{align}
The Hamiltonian $\hat H$ (\ref{eq:H-sigma}) has quantized eigenvalues $E_{S,n}$. Their 
 dependence  on the quantum number $n$  for each irreducible subspace $\ell=1,1-2/N,1-4/N,\ldots $ can be obtained using the discrete WKB approach \cite{Garg-wkb,Semerjian-wkb}. 
 
The eigenfunction $C_m^{\ell}$ in the classically forbidden region of $m$ is obtained from the analysis of an auxiliary classical one-dimensional system
with energy $e$, coordinate $m$, and imaginary \lq\lq momentum" $\hat p\rightarrow -i p$ that 
 obeys the Hamilton-Jacobi equation \cite{Semerjian-wkb}
 \begin{equation}
\varepsilon_\ell(m,p)\equiv -\Gamma \sqrt{\ell^2-m^2}\cosh p- g(m) =e\,. \label{eq:Hcl}
\end{equation}
In the classically-forbidden region  $\cosh p\geq 1$, leading to the condition
\begin{equation}
 -\big[e+g(m)\big] > \Gamma \sqrt{\ell^2-m^2} >0 \,.\label{eq:fr}
 \end{equation}
From Eq.~(\ref{eq:Hcl}) we obtain the momentum 
\begin{equation}
p\equiv p_\ell(m, e)=\arcsinh \frac{\sqrt{\big(e+g(m)\big)^2-\Gamma^2(\ell^2-m^2)}}{\Gamma\sqrt{\ell^2-m^2}}\;.\label{eq:kr}
\end{equation}
At the  boundaries of the forbidden region $m=a_i$ ($i$=0,1) and $p_\ell(a_i,e)=0$. These points can be obtained from the solution of the transcendental equation $U_\ell(m)=e$ where for the effective potential 
\begin{equation}
U_\ell(m)\equiv \varepsilon_\ell(m,0)=-\Gamma \sqrt{\ell^2-m^2}- g(m).   \label{eq:Ueff}
\end{equation}
Figure~\ref{fig:Veff} shows plots of the effective potential $U_\ell(m)$ for the Curie-Weiss model where interaction energy density $-g(m)$ takes a form 
\begin{equation}
-g(m)=-\frac{1}{2} m^2-h m\label{eq:CW}\,.
\end{equation}

For values of total momentum $\ell$ and  energy $e$ where metastability exists,  the eigenfunction under the barrier takes a standard  WKB form \cite{Garg-wkb}
\[C_m^\ell  \propto \frac{1}{\sqrt{\partial p_\ell(m,e)/\partial e}} \exp\left(- \int^{m}_{a_0(e,\ell)} dm^\prime  p_\ell(m^\prime,e)\right )\;.\]
Then, to exponential accuracy in $N$, the amplitude of the tunneling decay of the metastable state equals to
\begin{gather}
C_{\rm tunn}(e,\ell) \propto  \exp\left(-\frac{N}{2}\,  S_\ell(e)  \right), \label{eq:Ctunn}
\end{gather}
where $S_\ell(e) $ is the mechanical action under the barrier 
\begin{equation}
S_\ell(e) =\int_{a_0(e,\ell)}^{a_1(e,\ell )}  p_\ell(m,e)\,  dm\,, \label{eq:a}
\end{equation}
and  $m=a_i$ are
boundaries of a classically-forbidden region (cf. Fig.~\ref{fig:Veff}).

In the WKB approach one can also compute the action $S_\ell(e)$ by introducing the instanton  trajectory in imaginary time $t=-i\tau$ for a system with coordinate $m$, momentum $i p_\ell $, and velocity $i v_\ell $. Since we have introduced rescaled variables (\ref{eq:rescaled}), we also have to rescale the time $\tau$ in order to preserve the canonical relations
\begin{equation}
s=2\tau, \quad v_\ell=\frac{dm_{\scriptscriptstyle\rm WKB}}{ds} =-\left(\frac{\partial p_\ell}{\partial e}\right )^{-1}\;. \label{eq:resc}
\end{equation}
Using this relation we obtain the instanton trajectory $m=m_{\scriptscriptstyle\rm WKB}(s)$ in imaginary time $s$
\begin{align}
\frac{d m_{\scriptscriptstyle\rm WKB}}{ds} &=v_\ell (e,m_{\scriptscriptstyle\rm WKB}) \label{eq:nuH1} \\
& = \sqrt{\big[e+g(m_{\scriptscriptstyle\rm WKB})\big]^2-\Gamma^2(\ell^2-m^2_{\scriptscriptstyle\rm WKB})}\;.\nonumber
\end{align}
Then the action is $S_\ell(e)$=$\int_{0}^{s_0}p_\ell v_\ell\,ds$ where $s_0$ is the duration of motion under the barrier
\begin{equation}
s_0(e,\ell)=-\frac{\partial S_\ell}{\partial e}=\int_{a_0(e,\ell)}^{a_1(e,\ell )}   \frac{dm}{v_\ell(e,m)}\;.\label{eq:dade}
\end{equation}

\subsection{Transition rate for thermally assisted tunneling}

Thermally assisted quantum tunneling is  relevant to the study of QMC, which is always implemented at finite temperatures. The rate of the thermally-assisted tunneling $W_{\rm tunn}$ can be written in the form (\ref{eq:F})
\begin{align}
W_{\rm tunn}&=\sum_{\ell,n} W_{\ell,n} \frac{\Omega_{\ell} \,e^{-\beta N e_{\ell,n}}}{Z_0} \\Z_0&=\sum_{\ell,n}\Omega_{\ell}\,e^{-\beta N e_{\ell,n}}\label{eq:Ws}\;.
\end{align}
Here $W_{\ell,n}\propto  |C_{\rm tunn}(e_{\ell,n},\ell)|^2$  is the tunneling decay rate of the states with the set of quantum numbers $\ell$ and $n$, and $\Omega_{\ell}$ gives the number of these states. In accordance with the general prescription  (\ref{eq:F}), each term in (\ref{eq:Ws}) is proportional to the probability of thermal activation  to the states with the energies $e_{\ell,n}$ that subsequently undergoes a tunneling decay.

In the limit of large $N$ one can replace the summations $\sum_{\ell,n}$ with $\int d\ell\int dn $ and take the integrals using the method of steepest descent. To the leading order
 one simply needs to  maximize the logarithm of the integrand $2\ln [C_{\rm tunn}(e,\ell)]+N Q_\ell-N \beta e$ with respect to $\ell$ and $e$ [here we used Eq.~(\ref{eq:Q})]. In this way, one replaces the sum of the thermally assisted  tunneling decays rates  over many channels by the rate for the most probably channel (optimal fluctuation). 

We refer to \cite{kechedzhi_open-system_2016}  for details of this analysis and simply provide a final result
\begin{equation}
W_{\rm tunn}=B_{\rm tunn} \exp(-N \alpha),\quad \alpha=\beta ({\mathfrak F}-{\mathfrak F}_0)\,,\label{eq:Wta}
\end{equation}
where $B_{\rm tunn}$ is a prefactor and ${\mathfrak F}-{\mathfrak F}_0$ is the difference of the effective \lq\lq free energies" for the thermally assisted tunneling transition. Here,
\begin{equation}
\beta\,{\mathfrak F}=\min_{e,\,\ell}\big(\beta e + S_\ell(e)-Q(\ell)\big)\,,\label{eq:Amin}
\end{equation}
and ${\mathfrak F}_0=-\lim_{N\rightarrow\infty}(\ln Z_0)/(N \beta)$  is a  free energy per spin of the metastable state
\begin{equation}
\beta \, {\mathfrak F}_0=\min_{\ell}\big(\beta e_0(\ell)-Q(\ell)\big)\,.\label{eq:R}
\end{equation}
Here, $e_0(\ell)=U(a_0)$ is the energy of the metastable minimum. The optimal value of $\ell$ in (\ref{eq:R}) minimizes the system free energy over the  irreducible subspaces associated with the  minimum
\begin{equation}
\ell_0={\rm argmin} \big(\beta e_0(\ell)-Q(\ell)\big)\,.\label{eq:l0}
\end{equation}

 The extremal conditions for Eq.~(\ref{eq:Amin}) read as
\begin{gather}
\beta =-\frac{\partial S_\ell(e)}{\partial e}\;, \\[3pt]
\frac{\partial Q_\ell}{\partial \ell} =\frac{\partial S_\ell}{\partial \ell}\,.
\end{gather}
Using  (\ref{eq:dade}) and (\ref{eq:Q}) we get
\begin{equation}
\ell = \tanh\left|\frac{\partial S_\ell(e)}{\partial \ell}\right |,\quad s_0(e,\ell)=\beta\,,\label{eq:opt}
\end{equation}
\begin{equation}
\left |\frac{\partial S_{\ell}(e)}{\partial \ell}\right |=\ell \int_{0}^{\beta}\frac{|e+g(m)|}{\ell^2-m^2}d\tau\label{eq:al}\;.
\end{equation}
As expected, the most probable channel for decay corresponds to the instanton $m_{\scriptscriptstyle\rm WKB}(s)$ of period $\beta$.

\section{Kramers escape rate for QMC tunneling simulations\label{sec:KramersQMC}}

\subsection{Simulating tunneling via Quantum Monte Carlo dynamics}

The  partition function $\m Z$ of the model \eqref{eq:H-sigma} at
inverse temperature $\beta$ can be obtained   by using the Suzuki-Trotter
formula  to map a quantum problem   to a  classical one  with one
additional (imaginary-time)  dimension $\tau\in (0,\beta)$ \cite{suzuki1976relationship}.  In the limit of infinite number of Trotter slices,  $\m Z$ is given by an integral over the  array of spin  paths    $\underline{\sigma}(\tau)=\{\sigma_j(\tau)\}_{j=1}^{N}$
where each path $\sigma_j(\tau)=\pm 1$ is  periodical along the imaginary-time axis   $\sigma_j(0)=\sigma_j(\beta)$ and parametrized by the locations of points (\lq\lq kinks") at the axis where the sign of $\sigma_j(\tau)$ changes \cite{prokof1998exact,*rieger1999application}.
The  Gibbs probability distribution over paths   $\underline{\sigma}(\tau)$ has the form 
\begin{align}
\m
  P_G[\underline{\sigma}(\tau)]&=\m Z^{-1} \prod_{i=1}^{N}\Gamma^{\kappa[\sigma_{j}(\tau)]}e^{-N  \int_{0}^{\beta}g [m(\tau)  ] d\tau},\label{eq:cF}
\end{align}
\noindent
where $m(\tau)=N^{-1} \sum_{j=1}^{N}\sigma_j(\tau)$ and the function
$\kappa[\sigma_j(\tau)]$ equals to the number of kinks in a  given path $\sigma_j(\tau)$.

QMC  samples from the   Gibbs distribution  (\ref{eq:cF}) using the Metropolis-Hastings algorithm by implementing a series of stochastic updates  of  the state vector of individual spin paths    $\underline{\sigma}(\tau)$.  At each QMC step, a transition  is proposed from a current system state $\underline{\sigma}(\tau)$ to a new candidate state
 $ \underline{\sigma}^{\prime}(\tau)$ based on a certain stochastic update rule [often using cluster update methods  for paths $\sigma_j(\tau)$ \cite{prokof1998exact,*rieger1999application}]. The  rule is  specific for   a 
 given implementation of the algorithm, however, the {\it ratio} of the acceptance and rejection probabilities of the transition  proposal $p_{\underline{\sigma}(\tau), \underline{\sigma}^{\prime}(\tau)}/(1-p_{\underline{\sigma}(\tau), \underline{\sigma}^{\prime}(\tau)})$  is the same for any implementation and equal to  
 $ P_G[ \underline{\sigma}^{\prime}(\tau)]/ P_G[\underline{\sigma}(\tau)]$. We assume that the   QMC updates  of  $\underline{\sigma}(\tau)$ occur at  the sequence of random instants of time $(t_1,t_2,\ldots )$  sampled from  a  Poisson process with a constant rate $\lambda$. Then, the  stochastic time evolution  of the state vector $\underline{\sigma}(\tau,t)$ corresponds to   the master equation for the probability distribution  function  $\m P_{\underline{\sigma}(\tau)}(t)$:
 \begin{equation}
 \frac{\partial P_{\underline{\sigma}(\tau)}}{\partial t}=\sum_{\underline{\sigma}^{\prime}(\tau)} w_{\underline{\sigma}^{\prime}(\tau), \underline{\sigma}(\tau)}\m P_{\underline{\sigma}^{\prime}(\tau)}-w_{\underline{\sigma}(\tau), \underline{\sigma}^{\prime}(\tau)}\m P_{\underline{\sigma}(\tau)}\label{eq:master}\;.
 \end{equation}
 Here,  $w_{\underline{\sigma}^{\prime}(\tau), \underline{\sigma}(\tau)}\propto \lambda p_{\underline{\sigma}^{\prime}(\tau), \underline{\sigma}(\tau)}$ are  transition probabilities that depend explicitly of the stochastic update rule. Based on the above, they  must  obey the detailed-balance conditions
 \begin{equation}
\frac{w_{\underline{\sigma}(\tau), \underline{\sigma}^{\prime}(\tau)}}{w_{\underline{\sigma}^{\prime}(\tau), \underline{\sigma}(\tau)}}=\frac{P_G[ \underline{\sigma}^{\prime}(\tau)]}{P_G[ \underline{\sigma}(\tau)]}\label{eq:DB}
\end{equation}
that guarantee that Gibbs distribution (\ref{eq:cF}) is the stationary (long-time) solution  of (\ref{eq:master}).

Because of the mean-field  character of the model (\ref{eq:H-sigma}) it is possible to
obtain in a closed form a  Gibbs probability measure for the magnetization per
spin order parameter $m(\tau)$:
\begin{equation}
P_G[m(\tau)]=\frac{e^{-N \beta \m F[m(\tau)]}}{Z_{0}},\, \int D m(\tau) P_G[m(\tau)]=1\label{eq:Z00}
\end{equation}
\noindent
and the free-energy functional $\m F$ has the form \cite{jorg2010energy}
\begin{multline}
\m F[m(\tau)] =\frac{1}{\beta} \int_{0}^{\beta}[m(\tau) g^\prime(m(\tau))-g (m(\tau))] d\tau \\
- \frac 1 \beta \ln \Lambda[g^\prime(m(\tau))]\label{eq:Fm}\;.
\end{multline}
Here, $-g(m)$ is the mean-field interaction energy density (\ref{eq:H-sigma}), $g^\prime(m)=dg/dm$, and 
the functional $\Lambda[\lambda(\tau)]$ equals
\begin{gather}
\Lambda[\lambda(\tau)] = {\rm Tr}K^{\beta,0}[{\bf B}(\tau)],\label{eq:Lambda}\\
K^{\tau_2,\tau_1}[{\bf B}(\tau)] ={\rm T_{+}}e^{-\int_{\tau_1}^{\tau_2} d\tau  H_0(\tau)}\,,\label{eq:propagator}\\
H_0(\tau) =-\mathbf{B}(\tau)\cdot \bm{\sigma},\quad \mathbf B(\tau)=(\Gamma,0,\lambda(\tau))\label{eq:H0}\;,
\end{gather}
where $\bm\sigma=(\sigma_x,\sigma_y,\sigma_z)$ is the vector of Pauli
matrices. The propagator $K^{\tau_2,\tau_1}[{\bf B}(\tau)]$ corresponds to a spin-1/2 particle evolving in imaginary time under the action  of the magnetic field $\mathbf B(\tau)$. 

If we consider the order parameter as static and remove the ``time''
indices, we obtain the free energy function 
\begin{multline}\label{eq:static_free_energy}
F(m)= m\,g^\prime(m)-g(m)\\-\frac 1 \beta \ln \Big(
2\cosh(\beta\sqrt{(g^\prime(m))^2+\Gamma^2})\Big)\;.\end{multline}
We denote the minima of the free energy by $m = m_i$, $F_i = F(m_i)$. Here, the index
$i=0$ ($i=1$) corresponds to the local (global) minimum of $F(m)$.

Quantum tunneling  can be simulated with QMC  using a general approach that was first considered in a QA context \cite{Santoro29032002,PhysRevB.66.094203,*martonak_quantum_2004,*battaglia_optimization_2005,*santoro_optimization_2006,boixo2014,Heim12032015}.
At $t=0$ the QMC state vector $\underline{\sigma}(\tau,0)$ is 
initiated in the vicinity of the metastable state basin  by uniformly sampling from the spin configurations that satisfy the conditions $ N^{-1}
\sum_{j=1}^{N}\sigma_j(\tau,0) \simeq m_0$.  Every time when  the state  vector
$\underline{\sigma}(\tau,t)$ arrives at  the vicinity of the global minimum $N^{-1} \sum_{j=1}^{N}\sigma_j(\tau,t_f) \simeq m_1$, the QMC process is terminated. Repeating this experiment many times, one can determine the average escape time of the QMC process from the metastable  states $m_0$, which is proportional to the inverse of the escape rate $W_{\rm QMC}$~\cite{isakov_understanding_2016}. The numerical studies of $W_{\rm QMC}$ are described in Sec.~\ref{sec:QMCnum}.

\begin{figure}[t]
  \includegraphics[width=3.0in]{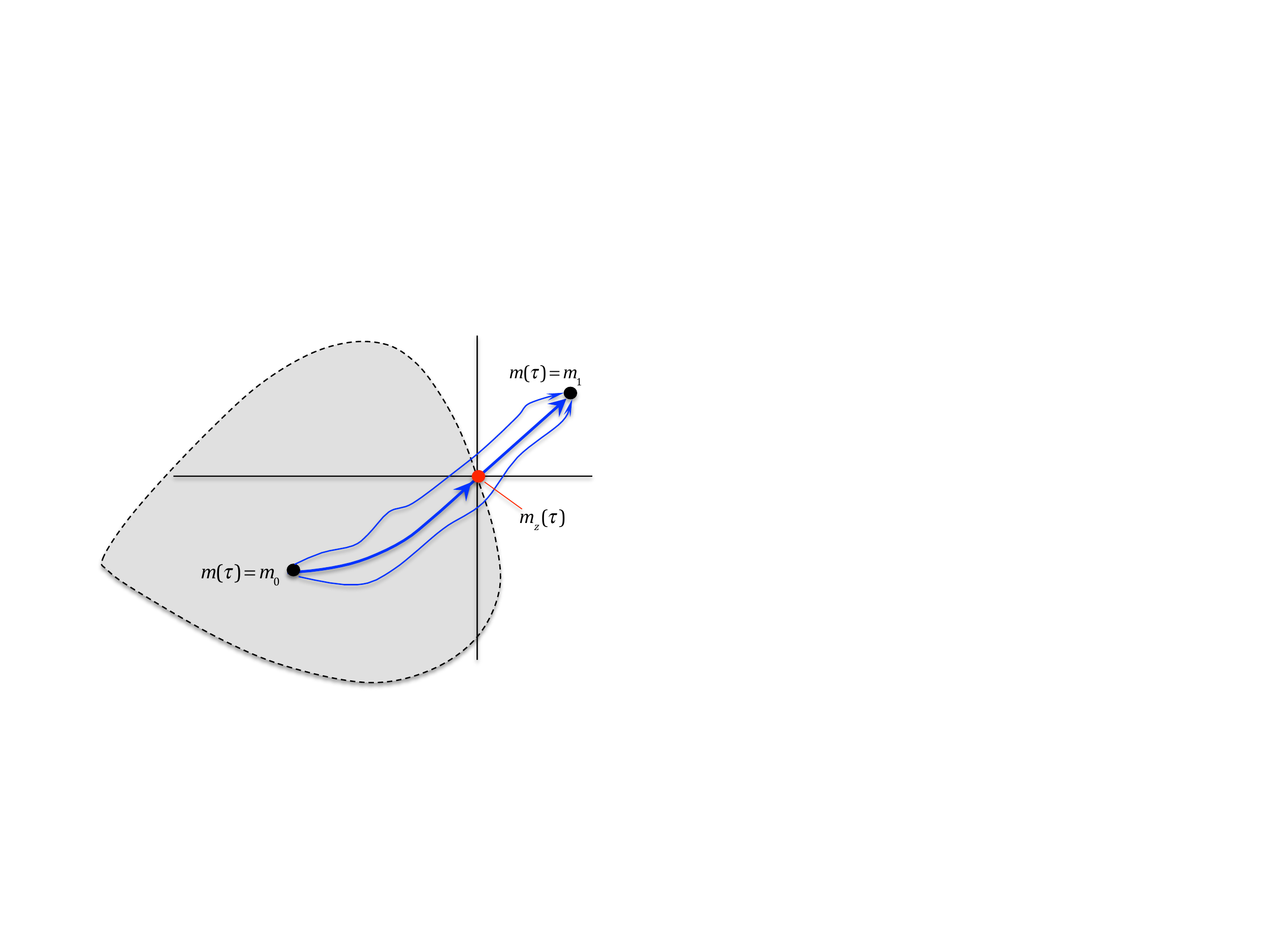}
    \caption{Schematic illustration of the stochastic Kramers escape paths  $m(\tau,t)$ depicted with blue lines. The most probable escape path is shown with a solid blue line. It starts near   $m_0$ and arrives  at    $m_1$ crossing the vicinity of saddle point $m_z(\tau)$ shown with a red dot. The dashed line depicts the boundary of the region where $ \m F[m(\tau)]\leq \m F[m_z(\tau)]$. Inside this region $\m F[m(\tau)]-\m F[m_z(\tau)]\gg T=\hbar/\beta$.} \label{fig:Kramers}
\end{figure}

It is convenient to study the stochastic trajectories $\underline{\sigma}(\tau,t)$ by
inspecting their projections $m(\tau, t)$ onto the continuous functional space defined in~\eqref{eq:Fm} (see Fig.~\ref{fig:Kramers}). At low enough temperatures  the distribution $P[m(\tau),t]$ corresponding to the QMC dynamics  (\ref{eq:master}) quickly relaxes toward quasi-equilibrium (\ref{eq:cF}) sharply localized in the vicinity of  $m_0$.  This relaxation process occurs with the   rate  $\gamma\gg W_{\rm QMC}$. The trajectory $m(\tau,t)$ spends a long time ($\sim W_{\rm QMC}^{-1}$) near the metastable state $m_0$. Occasionally, a large fluctuation occurs corresponding to the escape event where the path $m(\tau,t)$ moves away from $m_0$ and arrives eventually at  the vicinity of  $m_1$ (see Fig.~\ref{fig:Kramers}). During the escape event the system initially moves uphill when the free energy $\m F[m(\tau,t)]$ is increasing until it reaches the saddle point  of the functional $\m F$ to be denoted as $m_z(\tau)$ that satisfies the equation
\begin{equation}
\left . \frac{\delta \m F[m(\tau)]}{\delta m(\tau)}\right |_{m(\tau)=m_z(\tau)}=0, \quad m_z(0)=m_z(\beta)\;.\label{eq:var}
\end{equation}
Near the saddle point the escape path $m(\tau,t)$ slows down, because the variation of $\m F[m(\tau)]$ is small. Then it goes downhill almost deterministically, so that the free energy $\m F[m(\tau,t)]$ is decreasing until the state $m_1$ is reached. The quasi-stationary statistical distribution over $m(\tau)$ has the Gibbs form $P_G[m(\tau)]$ everywhere in the domain of the local minimum except in the small vicinity of the saddle point $| \m F[m(\tau)]-\m F[m_z(\tau)]|\lesssim \beta^{-1}$, where deviations from $P_G[m(\tau)]$ allow for the probability current flow away from the metastable state. This area lies inside the domain marked with dashed line in Fig.~\ref{fig:Kramers}. The probability for the path $m(\tau,t)$ to reach the vicinity of saddle point  that lies inside this  domain   is $P_G[m(\tau)]\propto \exp\{-F[m(\tau)]-F[m_0]\}$. This is precisely the Boltzmann factor that determines the transition rate exponent in  Kramers' theory of escape \cite{langer_statistical_1969,dykman1979theory,kamenev2011field,hanggi1990reaction}
\begin{equation}
\hspace{-0.1in} W_{\rm QMC} =B_{\rm QMC}\, e^{-\beta N  \Delta \m F },\quad  \Delta \m F=\m F[m_z(\tau)]-F(m_0)\;,\label{eq:WQMC}
\end{equation}
where $ \Delta \m F\gg \beta^{-1}$ and $B_{\rm QMC}$ is a prefactor (polynomial in $N$ ) that depends on the specific path update rule used in QMC. 


The connection between the saddle points of the free energy functional in a classical one-dimensional field theory and the instantons in the corresponding quantum mechanical tunneling problem was first established using the path-integral formalism in \cite{langer_theory_1967,langer_statistical_1969} for the case of a particle in a potential.
To apply the same argument to our mean-field quantum spin problem we use the partition function $\m Z_0$ associated with the Hamiltonian (\ref{eq:H-sigma}) in a form of the path integral normalization factor for  
$P_G[m(\tau)]$ (\ref{eq:Z00}). Then we express the tunneling decay rate in a standard form (\ref{eq:F}) in terms of the imaginary part of the partition function. For a large number of spins $N\gg 1$ this path-integral can be calculated using the saddle-point method within the instantonic calculus, where the free-energy functional $\m F[m(\tau)]$ plays the role of the mechanical action. This  gives 
\begin{equation}
W_{\rm tunn}=B_{\rm tunn} \exp(-\beta N  \Delta \m F)\,,\label{eq:WtunnQMC}
\end{equation}
 where $ \Delta \m F$ is determined by (\ref{eq:var}) and (\ref{eq:WQMC}). The dominant  contribution to the path integral is given by the instanton $m_z(\tau)$. We will show below  that the exponential scaling of $W_{\rm tunn}$ and $W_{\rm QMC}$ with $N$ is the same. The prefactors $B_{\rm QMC}$ and $B_{\rm turn}$ are expected to be different.   $B_{\rm tunn}$ can be expressed entirely in terms of the free-energy functional $F[m(\tau)]$  while $B_{\rm QMC}$ depends on the specific path update rule used in QMC.

To complete our analysis, we would like to relate our findings to the WKB analysis of thermally assisted tunneling presented in Sec.~\ref{sec:WKB}.  We will  now show in an explicit analytical form that  the exponential scaling coefficient $\beta \Delta \m F$ of the QMC transition rate  with $N$ and the  QMC saddle-point solution  $m_z(\tau)$ are identical to, respectively, the exponential scaling coefficient 
$\alpha$ [Eq.~(\ref{eq:Wta})] of the WKB tunneling rate  with $N$ and the WKB instanton tunneling path $m_{\rm WKB}(\tau)$ [Eq.~(\ref{eq:nuH1})].

Such analysis is of interest because unlike the extrema of the free-energy functional (\ref{eq:static_free_energy}) that obey the \lq\lq static" condition $m(\tau)=m_{0,1}$, the saddle-point (instanton)  trajectories $m_z(\tau)$  are time dependent.  They also appear  in the context of the QA in mean-field spin models  \cite{boulatov2003quantum,jorg2010energy,seoane2012many,Semerjian-wkb} of the type given in (\ref{eq:H-sigma}).
 
\section{Comparison of WKB and QMC transition rates \label{sec:comp}}

Using the expression (\ref{eq:Fm}) for $\m F$, we can re-write Eq.~(\ref{eq:var}) in the form  of the following two equations:
\begin{equation}
m_z(\tau)=\frac{\delta }{\delta \lambda(\tau)}\ln \Lambda[\lambda(\tau)],\quad \lambda(\tau)= g^\prime[m_z(\tau)]\label{eq:dm-var}\;,
\end{equation}
where the functional $\Lambda[\lambda(\tau)]$ is defined in  (\ref{eq:Lambda})--(\ref{eq:H0}).

We now introduce a vector function in imaginary time ${\bf
  m}(\tau)$=$(m_x(\tau),m_y(\tau),m_z(\tau))$ corresponding to expectation values of
the operator ${\bm \sigma}$ for the spin-$\frac{1}{2}$  particle defined as 
\begin{align}
  {\bf  m}(\tau) &= \frac{{\rm Tr} \left (K^{\beta,\tau}  {\bm \sigma}
  K^{\tau,0}\right )}{{\rm Tr}\,K^{\beta,0} } \nonumber \\ &= \frac{\delta }{\delta {\bf B}(\tau)} \ln {\rm Tr}
  K^{\beta,0}[{\bf B} (\tau)]\label{eq:dm-vec}\;,
\end{align}
where ${\bf B} (\tau)$ is defined in Eq.~\eqref{eq:H0}.
We recognize that Eq.~\eqref{eq:dm-var} corresponds to the
$z$-component of the vectorial equation~\eqref{eq:dm-vec}.


Differentiating this equation with respect to $\tau$ and using (\ref{eq:H0}), we obtain
\begin{equation}
\frac{d{\bf m}}{d\tau} =\frac{{\rm Tr} \left (K^{\beta,\tau} [H_0(\tau), \bm{\sigma}] K^{\tau,0}\right )}{{\rm Tr}\,K^{\beta,0} }\;.\label{eq:mvec}
\end{equation}
One can re-write this equation in the following form
\begin{equation}
\frac{d{\bf m}}{d\tau}=-2 i \frac{\partial {\mathcal H_0}({\bf
    m})}{\partial {\bf m}} \times {\bf m}\,, \label{eq:dmvec}
\end{equation} 
where
\begin{equation}
 {\mathcal H_0}({\bf m}) = -\Gamma m_x-g(m_z)\label{eq:H0c}\;.
 \end{equation}
and we observe that ${\bf B}(\tau) = -{\partial {\mathcal H_0}[{\bf m}(\tau)]}/{\partial {\bf m}(\tau)}$.
We have to solve this equation with the periodic boundary condition ${\bf m}(0)={\bf m}(\beta)$ [cf.~(\ref{eq:var})].

Equation~(\ref{eq:dmvec}) allows for two integrals of motion
\begin{gather}
{\cal H}_{0}({\bf m})=e,\label{eq:int1}\\
{\bf m}(\tau)\cdot {\bf m}(\tau)=\ell^2 \label{eq:int2}
\end{gather}
where  ${\bf m}(\tau)\cdot {\bf m}(\tau)\equiv m_x^2(\tau)+m_y^2(\tau)+m_z^2(\tau)$.
Then, the solution of (\ref{eq:dmvec})  can be written in the form:
\begin{gather}
m_x =\sqrt{\ell^2-m_z^2}\cosh p_\ell(m,e)\;,\label{eq:mx}\\
m_y = -i \sqrt{\ell^2-m_z^2} \sinh  p_\ell(m,e)\;,\\
\frac{d m_z}{d\tau}  =2v_\ell(e,m)\label{eq:mz}\;,
\end{gather}
where $p_\ell$ and $v_\ell$ are given in Eqs.~(\ref{eq:kr}) and (\ref{eq:nuH1}), respectively. The equation for $m_z(\tau)$ is identical to that for the WKB instanton trajectory  $m_{\scriptscriptstyle\rm WKB}(\tau)$ defined in (\ref{eq:nuH1}).

Because $\ell$ is a constant of motion, its value can be determined at $\tau=0$:
\begin{align}
 \ell = \frac{1}{\tr K^{\beta, 0}} \sqrt{\sum_{j=x,y,z}\big(\tr(K^{\beta, 0}\,\sigma_j)\big)^2  }\;.\label{eq:l_self}
\end{align}
Since the propagator $K^{\beta, 0}$ depends on $\ell$, Eq.~({\ref{eq:l_self}) is a self-consistent condition.

\subsection{The equivalence of WKB and QMC instanton trajectories \label{sec:double}}

We have  shown above that the saddle point of QMC satisfies the same differential equations (\ref{eq:mz}) and (\ref{eq:nuH1}) as the WKB instanton. What remains to be shown is that the optimal value of $\ell$ for the WKB approach coincides with that given by the self-consistent condition of QMC (\ref{eq:l_self}). In the WKB approach, the optimal value of $\ell$ can be determined by the conditions~(\ref{eq:opt}) and (\ref{eq:al}). In this section, we express the self-consistent condition for QMC in the same form as the corresponding condition in WKB to demonstrate their equivalence. 

It is useful to introduce a replica qubit, which allows us to write the self-consistent condition~(\ref{eq:l_self}}) as
\begin{align}\label{eq:l_self_d}
\begin{split}
 \ell^2 \ssp(\tr K^\beta)^2 &= \tr\Big(K^\beta\otimes K^\beta\!\sum_{j=x,y,z}\sigma_j\otimes\sigma_j \Big)\\
 &= \tr\Big(K^\beta\otimes K^\beta \big(P_S-3P_A\big) \Big)\;,
 \end{split}
\end{align}
where $K^\beta\equiv K^{\beta, 0}$ and $P_S$ ($P_A$) is the projector onto the symmetric (anti-symmetric) subspace of the two qubits. To analyze  the double propagator $K^\beta\otimes K^\beta$, we consider the Hamiltonian 
\[H_0^{(2)} = H_0\otimes I + I\otimes H_0,\quad H_0 =-\Gamma \sigma_x-\g'(\mm_z) \sigma_z \]
 where $H_0$ was already introduced in   Eq.~(\ref{eq:H0}).
With the Bell basis $\ket{\Phi^+} = \frac{1}{\sqrt{2}}(\ket{00}+ \ket{11})\,,\;
\ket{\Phi^-} = \frac{1}{\sqrt{2}} (\ket{00}- \ket{11})\,,\;
\ket{\Psi^+} = \frac{1}{\sqrt{2}} (\ket{01}+ \ket{10})\,,\; 
\ket{\Psi^-} = \frac{1}{\sqrt{2}} (\ket{01}- \ket{10})$, we have
\begin{align}
&-H_0^{(2)} \ket{\Phi^+} = 2 \Gamma \ket{\Psi^+} + 2 \g'(\mm_z) \ket{\Phi^-}\;,\label{eq:Phi+}\\
&-H_0^{(2)}\ket{\Phi^-} = 2 \g'(\mm_z) \ket{\Phi^+}\;,\label{eq:Phi-}\\
&-H_0^{(2)}\ket{\Psi^+} = 2 \Gamma \ket{\Phi^+}\;,\label{eq:Psi+}\\
&-H_0^{(2)}\ket{\Psi^-} = 0\;.
\end{align}
The anti-symmetric singlet state $\proj{\Psi^-}$ is a dark state, and the triplet states are closed under such evolution. As a consequence, the following identity always holds
\begin{align}\label{eq:anti-symmetric_identity}
\tr \big(K^\beta\otimes K^\beta P_A\big)  = \tr P_A = 1\;,
\end{align}
with which we have
\begin{align}
\big(\tr K^\beta\big)^2 - \sum_{j=x,y,z}\big(\tr (K^\beta\sigma_j)\big)^2  
&= 4\;.
\end{align}
Using the identity~(\ref{eq:anti-symmetric_identity}), the self-consistent condition~(\ref{eq:l_self_d}) can be simplified to 
\begin{align}
 \ell^2  
 &= 1- \frac{4}
 {\tr\big(K^\beta\otimes K^\beta P_S\big) + 1}\;.\label{eq:l_trace}
\end{align}
To solve $\ell$, we need to know the trace of the double propagator in the symmetric subspace $\tr\big(K^\beta\otimes K^\beta P_S\big)$.

Consider the time evolution of a state $\ket{\Xi(\tau)}$ in the symmetric subspace of the two qubits,
\begin{align}
\begin{split}
 \ket{\Xi(\tau)} &= K(\tau, 0)\otimes K(\tau, 0)\ket{\Xi(0)}\\
 &=-\xi_x (\tau)\ket{\Phi^-} -i \xi_y (\tau)\ket{\Phi^+} +  \xi_z(\tau)\ket{\Psi^+}\;,
 \end{split}
\end{align}
where $\xi_x, \xi_y$ and $\xi_z$ take real values.
Using Eqs.~(\ref{eq:Phi+})--(\ref{eq:Psi+}), we have the following differential equations for the coefficients:
\begin{equation}
\frac{d{\boldsymbol \xi}}{d\tau}=-2 i \frac{\partial {\mathcal H_0}({\bf
    m})}{\partial {\bf m}} \times {\boldsymbol \xi} \;,\label{eq:dxivec}
\end{equation} 
where $\bf m$ is determined by the instanton equations~(\ref{eq:mx})--(\ref{eq:mz}). Equation~(\ref{eq:dxivec}) is very similar to Eq.~(\ref{eq:dmvec}), except that it is linearized. Thus, a known solution to Eq.~(\ref{eq:dxivec}) is the instanton solution ${\boldsymbol \xi}(\tau) = {\bf m}(\tau)$. Because the instanton solution is periodic, ${\bf m}(0)$ is an eigenvector of the propagator $K^\beta\otimes K^\beta$ with eigenvalue $1$. To evaluate the trace, we still need to know the other two linearly independent solutions to Eq.~(\ref{eq:dxivec}).

We define the following symmetric bilinear form
\begin{align}
 \sB\big(\boldsymbol \xi(\tau), \boldsymbol \eta(\tau)\big) = \xi_x(\tau)\eta_x(\tau) +\xi_y(\tau)\eta_y(\tau) +\xi_z(\tau)\eta_z(\tau) \,,
\end{align}
where $\boldsymbol \xi $ and $\boldsymbol \eta $ are two solutions to Eq.~(\ref{eq:dxivec}). The bilinear form takes real values, and it is not an inner product (there is no complex conjugate on $\xi_y$). It is readily seen that the bilinear form is a constant of motion from Eq.~(\ref{eq:dxivec}), 
\begin{align}
 \sB\big(\boldsymbol \xi(\tau), \boldsymbol \eta(\tau)\big) = \sB(\boldsymbol \xi(0), \boldsymbol \eta(0))\,.
\end{align}

Let the other two solutions to Eq.~(\ref{eq:dxivec}) satisfy the initial conditions 
\begin{gather}
 \xi_x^{(\pm)}(0) = -\mm_z(0)\,,\quad \xi_z^{(\pm)}(0) = \mm_x(0)\,,\\
 \xi_y^{(\pm)}(0) = \pm i\sqrt{\mm_x^2(0) +\mm_z^2(0)}\,.
\end{gather}
Notice that
\begin{align}
 \sB\big(\boldsymbol \xi^{(\pm)}(0),\, \boldsymbol \xi^{(\pm)}(0)\big) = \sB\big(\boldsymbol \xi^{(\pm)}(0),\, {\bf m}(0)\big) = 0\;.\label{eq:pseudo_product}
\end{align}
Because all the bilinear forms in Eq.~(\ref{eq:pseudo_product}) are preserved at time $\beta$ and ${\bf m}(\beta)={\bf m}(0)$, $\boldsymbol\xi^{(\pm)}(\beta)$ must proportional to $\boldsymbol\xi^{(\pm)}(0)$ or $\boldsymbol\xi^{(\mp)}(0)$. The second possibility can be ruled out by noticing that the signs of $i\xi_y^{(\pm)}(\tau)$ cannot change during the evolution, because $i\xi_y^{(\pm)}(\tau)\neq 0$ for all $\tau \in [0,\,\beta]$. As a result, we have 
\begin{gather}
\boldsymbol\xi^{(\pm)}(\beta) = \kappa_\pm\, \boldsymbol\xi^{(\pm)}(0)\;,\label{eq:solution_pm}
\end{gather}
where proportional factors $\kappa_\pm$ are to be determined. Using Eq.~(\ref{eq:pseudo_product}), we have
\begin{align}
 -\big(\xi_y^2+\xi_z^2\big) \mm_x^2 = -\xi_y^2 \mm_y^2 +2i \xi_z \xi_y \mm_z \mm_y + \xi_z^2 \mm_z^2\;,
\end{align}
where we neglect the superscripts $(\pm)$ in $\xi$ to simplify the notations.  Introducing $\ratio_\pm = -i\xi_y^{(\pm)}/ \xi_z^{(\pm)}$, we have
\begin{align}
 (\ratio^2_\pm-1) \mm_x^2 = -\ratio_\pm^2 \mm_y^2 +2 i\ratio_\pm \mm_z \mm_y + \mm_z^2\;.
\end{align}
The solution to the above quadratic equation is
\begin{align}
 \ratio_\pm &= \frac{i\mm_z \mm_y\pm \ell\, \mm_x}{\ell^2-\mm_z^2}\;,\label{eq:ratio}
\end{align}
where we use the condition~(\ref{eq:int2}) to simplify things.  Putting the definition of  $\ratio_\pm$ into Eq.~(\ref{eq:dxivec}), we have
\begin{align}
 \dt \xi^{(\pm)}_z(\tau) = 2 \Gamma \ratio_\pm(\tau) \xi^{(\pm)}_z(\tau)\;,
\end{align}
which can be solved exactly
\begin{align}
 \xi^{(\pm)}_z(\tau) = \xi^{(\pm)}_z(0)\, e^{2 \Gamma\! \int_0^\tau \ratio_\pm(\tau')\, \dif \tau'}\;.\label{eq:xi_z}
\end{align}
The integral in the above equation can be evaluated using the solution Eq.~(\ref{eq:ratio}),
\begin{align}\label{eq:sI}
\begin{split}
 \Gamma\int_0^\beta \ratio_\pm(\tau)\, \dif \tau
 &= \pm \ell\int_0^\beta \frac{ \normb{\erg+ g(\mm_z)}}{\ell^2-\mm_z^2}\, \dif \tau \equiv \pm \mathcal I\;,
 \end{split}
\end{align}
where we also use the relations $\mm_x = \normb{\erg+ g(\mm_z)}/\Gamma$ and $i\mm_y = \dt \mm_z/2 \Gamma$.  
The eigenvalues $\kappa_\pm$ can thus be determined,
\begin{gather}
\kappa_\pm = e^{2 \Gamma\! \int_0^\beta \ratio_\pm(\tau)\, \dif \tau} = e^{\pm 2 \mathcal I} \,.\label{eq:kappa}
\end{gather}
With the three eigenvalues of the double propagator all solved, we have
\begin{align}\label{eq:trace_K}
 \tr(K^\beta\otimes K^\beta P_S\big) = 1 + 2\cosh \big(2\mathcal I\big)\;.
\end{align}
Putting the above result into Eq.~(\ref{eq:l_trace}), we have
\begin{align}\label{eq:self_consis}
 \ell =\sqrt{ \frac{\cosh\big(2\mathcal I\big)-1}{\cosh\big(2\mathcal I\big)+1} }= \tanh\mathcal I\;.
\end{align}
The integral $\mathcal I$ defined in Eq.~(\ref{eq:sI}) takes the same form as Eq.~(\ref{eq:al}), although there is a time scale difference of $2$ between the two.  Comparing Eqs.~(\ref{eq:opt}) and (\ref{eq:self_consis}), we conclude that the optimal value of $\ell$ in the WKB approach equals to the self-consistent solution of $\ell$ in the QMC approach. 

\subsection{The equivalence of the WKB and  QMC  scaling of transition rate with the number of spins}

Taking into account the expression  for  $\mathcal I = \tan^{-1}\ell$  defined in Eq.~(\ref{eq:sI}), after some transformations the WKB action~(\ref{eq:a}) for the extremal trajectory corresponding to the instanton solution (\ref{eq:mz}) can be re-written in the following form:
\begin{align}
\hspace{-0.098in} \action  = -\beta \erg -\ell\,\mathcal I + \int_{0}^{\beta} \left[ m_z(\tau) \lambda(\tau)-g(m_z(\tau))\right]\,\dif \tau \,.
  \label{eq:wkb_action}
\end{align}
where $\lambda(\tau)=g^{\prime}(m_z(\tau))$ is given in (\ref{eq:dm-var}).   Using Eq.~(\ref{eq:trace_K}), we have the trace of the QMC propagator,
\begin{align}
\hspace{-0.11in} (\tr K^\beta)^2 &= \tr(K^\beta\otimes K^\beta P_S\big) +1 = 2 + 2\cosh \big(2\mathcal I\big)
\end{align}
or equivalently, 
\begin{align}
 \tr K^\beta = 2\cosh \mathcal I\;.\label{eq:trace_K1}
\end{align}
With Eq.~(\ref{eq:self_consis}), the logarithm of $\tr K^\beta$ can be expressed as a function of $\ell$,
\begin{align}\label{eq:ln_trace_K}
 \ln \big(\tr K^\beta\big) =  \ln \frac{2}{\sqrt{1-\ell^2}} \;.
\end{align}
From   (\ref{eq:wkb_action}) and (\ref{eq:ln_trace_K}) the QMC free energy (\ref{eq:Fm}) takes the form
\begin{align}
\beta \,\m F[m_z(\tau)] &=\int_{0}^{\beta} \big(\lambda(\tau) \mm_z(\tau)-g[\mm_z(\tau)]\big) \, d\tau-\ln \big(\tr K^\beta\big)\nonumber\\
&= \action+\beta \erg+ \ell\,\tan^{-1}\ell - \ln \frac{2}{\sqrt{1-\ell^2}}\;.\label{eq:free_energy_instanton}
\end{align}
Using the following identity from the definition of the entropic factor $Q_\ell$ (\ref{eq:Q}),
\begin{align}
 \ln \frac{2}{\sqrt{1-\ell^2}} = Q_\ell + \ell \tan^{-1}\ell \;,
\end{align}
we finally obtain
\begin{equation}
\beta \,\m F[m_z(\tau)]= \action+\beta \erg -Q(\ell)\,.
\end{equation}
 After the minimization over $\ell$ and $e$   is performed,  this has  exactly the same form as the expression for the effective free-energy  in the exponent of the WKB transition rate  (\ref{eq:Amin}) 
 \[\min_{e,\ell}\m F[m_z(\tau)]=  \mathfrak{F}\,.\]
 
 To complete the comparison we note that  the static free energy per spin $\mathfrak{F}_0$ that appears in (\ref{eq:Wta}) equals to the free energy density $F(m_0)$ from (\ref{eq:WQMC}).  Indeed, those are merely  two different expressions for the free energy per spin of the metastable state $m=m_0$. Connecting them requires a    cumbersome but straightforward calculation given in Appendix.  With that we can establish that 
 \[\beta \Delta \m F=\alpha\,,\]
  [cf. (\ref{eq:Wta}) and (\ref{eq:WQMC})].  Thus, we have shown the equivalence of the exponential factors in WKB and QMC transition rate expressions.

\section{Numerical Results \label{sec:QMCnum}}

\begin{figure*}[ht]
\centering
\subfloat[Bias = 0]{
  \includegraphics[width=.4\textwidth]{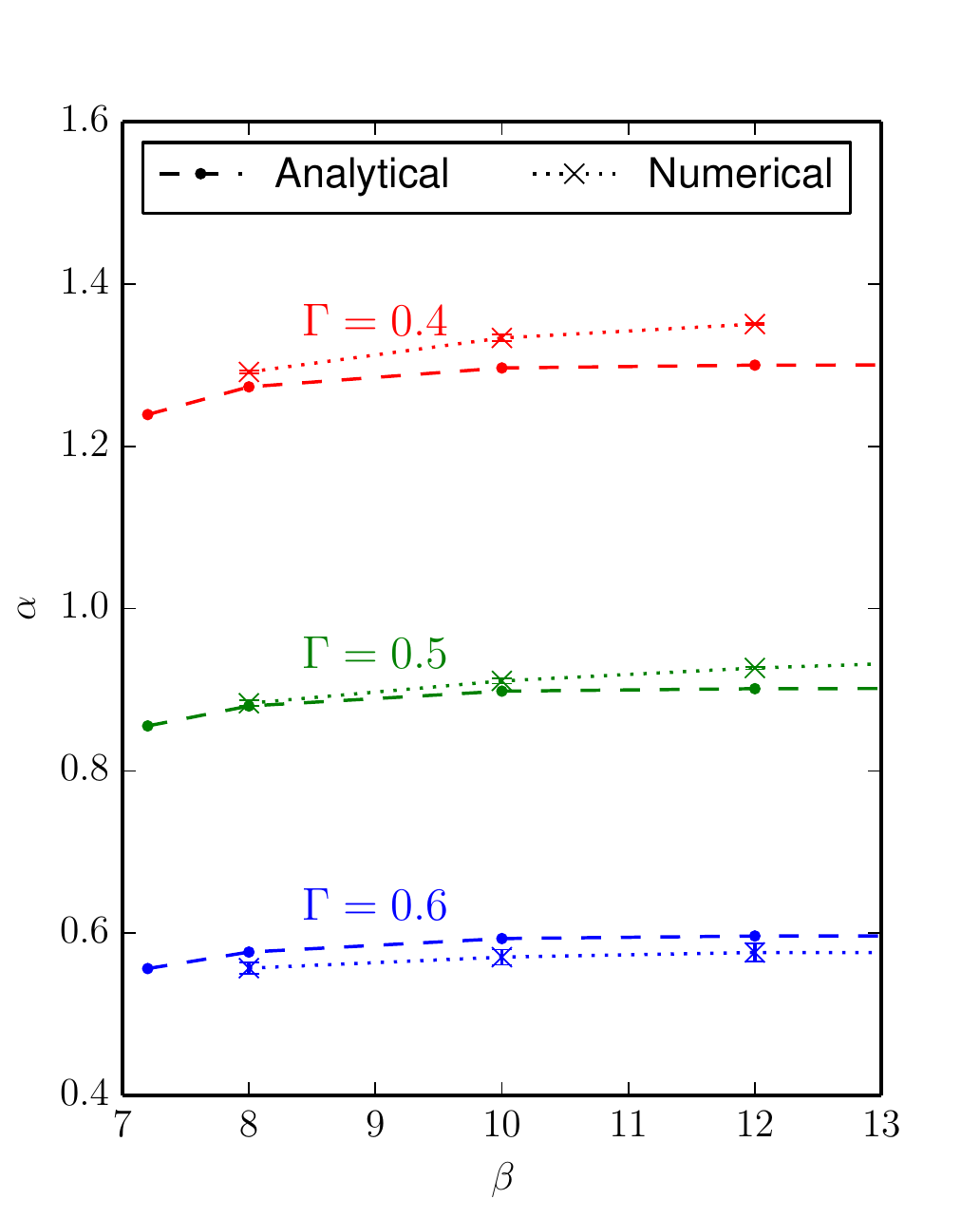}
\label{fig:qmc_comparison_bias0}
}
\subfloat[Bias = 0.1]{
  \includegraphics[width=.4\textwidth]{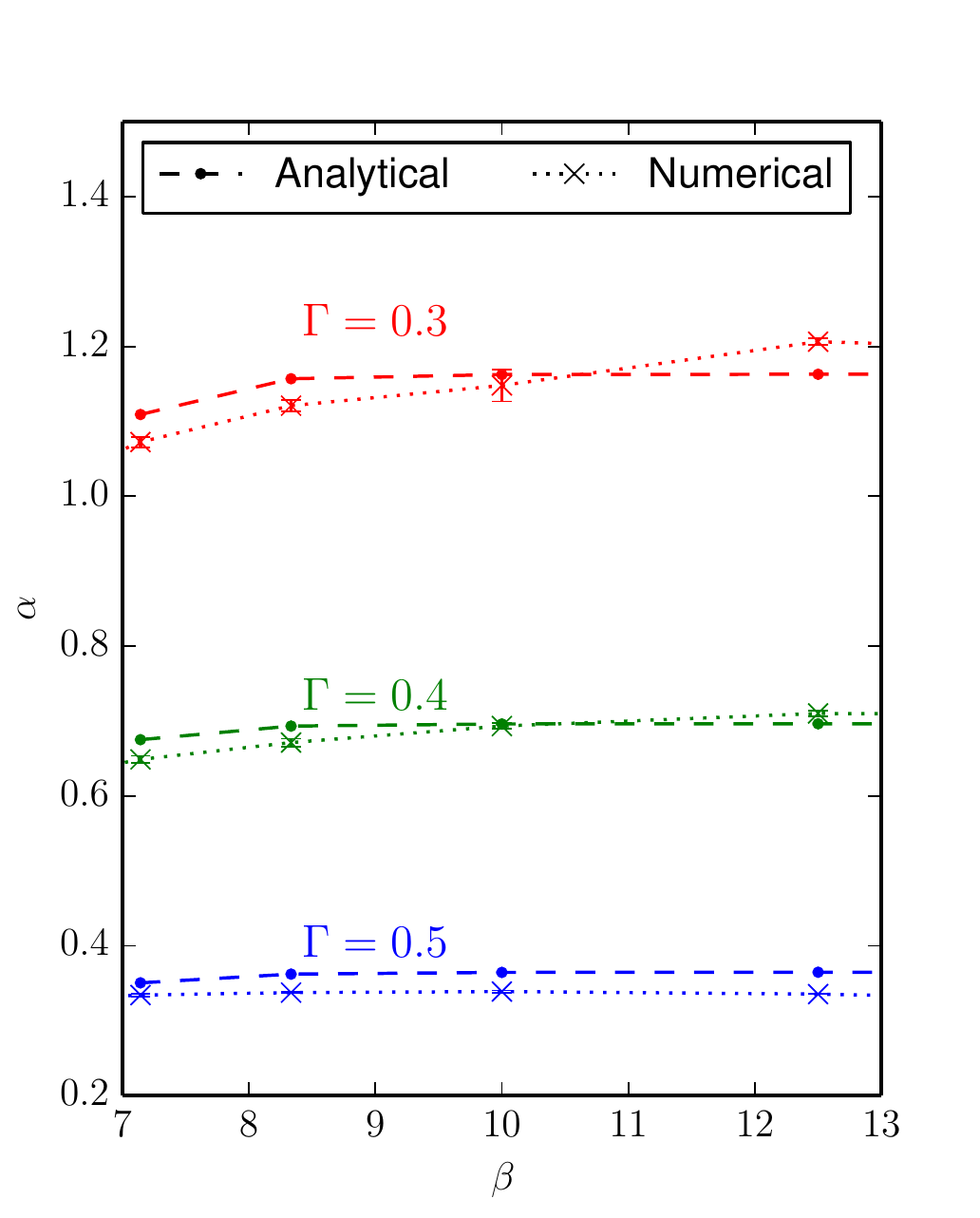}
  \label{fig:qmc_comparison_bias1}
}
\caption{Dependence on the inverse temperature $\beta$ of the  the scaling exponent $\alpha$ ($y$-axis) in the analytical expression for the tunneling decay rate (\ref{eq:Wta}) 
 and  the  scaling exponent for the  QMC escape rate obtained via continuous Path Integral Quantum Monte Carlo simulations. Both results are obtained for
 the  the Curie-Weiss model (\ref{eq:CW}) for  different values of transverse field $\Gamma$ and bias $h$. The $x$-axis
   corresponds to  different values for the inverse temperature $\beta$.
  (a) Different
   colors correspond to different values of the transverse field   $\Gamma \in \{0.4, 0.5, 0.6\}$ for 
  zero  bias  $h$ = 0.  (b) Different
   colors correspond to different values of 
   $\Gamma \in \{0.3, 0.4, 0.5\}$  with $h$ = $0.1$.
  In both figures the  $\cdot$
 symbols with dashed lines correspond to the analytical values while
 the $\times$ symbols with dotted lines correspond to numerical
 values. Error bars correspond to the numerical fitting of the
 exponent $\alpha$ in the numerical data.}
\end{figure*}

In this section we compare the analytical result for the tunneling escape rate based on Eq.~\eqref{eq:Wta} with the QMC escape rate obtained through numerical simulations performed with continuous-time path integral quantum Monte Carlo (QMC)~\cite{rieger1999application}. We study the quantum Curie-Weiss model with the Hamiltonian (\ref{eq:H-sigma}), where $g(m)$ is given in (\ref{eq:CW}). The corresponding effective potential $U_\ell(m)$ for $\ell = 2S/N=1$ is depicted in Fig.~\ref{fig:Veff}. 

We follow the same method as in Ref.~\cite{isakov_understanding_2016} to obtain the exponential scaling with $N$ in the numerical decay
rate in QMC simulations, $W_{\rm QMC}\propto \exp(-\alpha N)$. However, we extend the study \cite{isakov_understanding_2016} to the cases of nonzero biases and finite temperatures. We
initialize all spin worldlines in the neighborhood of the higher
(left) local minima of $U_\ell(m)$ by setting $\sigma_j(\tau) = -1$. We
measure the number of QMC sweeps (defined as one attempted update per
spin worldline) required to decay from the metastable state. This is
done by counting the number of sweeps until at least $25\%$ of the replicas reverse
their magnetization to $\sigma_j(\tau) = 1$~\footnote{The choice of 25\% is
  somewhat arbitrary. We find that waiting $25\%$ or
  $50\%$ of the replicas reverse magnetization results in only small
  differences in the fitting exponent $\alpha$. This difference is slightly larger for cases when the number of sweeps for a transition in QMC is small.}. We obtain the
average number of sweeps required for a given number of spins $N$ by
repeating this measurement a large number of times. This was done for
$N \in \{8, 10, 12, 14, 16\}$. The dependence of the number of sweeps on
$N$ is well fitted by the expression $\exp(\alpha N)/N$ (see
Ref.~\cite{isakov_understanding_2016}). We finally obtain the exponent
$\alpha$ from the exponential fit with $12 \le N \le
16$. In Figs.~\subref*{fig:qmc_comparison_bias0} and~\subref*{fig:qmc_comparison_bias1} the exponent $\alpha$ obtained from numerical QMC simulations is compared to that 
obtained from the analytical approach developed in the previous sections. 

For the analytical value, we solve numerically the  rate for thermally assisted tunneling from the expressions (\ref{eq:Fm}) and (\ref{eq:WtunnQMC}) (their equivalence to the WKB result (\ref{eq:Wta}) was shown in the previous section). We use the instanton solution Eq.~\eqref{eq:nuH1} for a particular $\ell = \ell_1$ to construct the 2-dimensional matrix $H_0(\tau)$ in Eq.~\eqref{eq:H0}, where $\lambda(\tau)$ is a function of the instanton solution. The 2-dimensional matrix $H_0(\tau)$ is then used to calculate the propagator Eq.~\eqref{eq:propagator}. Knowing the propagator allows one to calculate $\ell_2$ using Eq.~\eqref{eq:l_self}. If $\ell_1\neq \ell_2$, the original value $\ell_1$ is modified until it equals to the corresponding $\ell_2$; this value $\ell_1=\ell_2$ is the self-consistent solution $\ell^*$. The free energy $\m F[m_z(\tau)]$ of the instanton trajectory with $\ell^*$ (the saddle point) can thus be calculated using Eq.~\eqref{eq:Fm}. The static free energy for the metastable state $F(m_0)$ is calculated by minimizing $F(m)$ in Eq.~\eqref{eq:static_free_energy} over $m$. It can be shown analytically that the extreme points of Eq.~\eqref{eq:static_free_energy} equal to those of the effective potential (after optimizing over $\ell$). The escape rate $\beta(\m F[m_z(\tau)]-F(m_0))$ given in (\ref{eq:Fm}), (\ref{eq:WtunnQMC})  is proportional to the difference of the instanton free energy and the static free energy corresponding to the local minimum.

\section{Problems with small and narrow barrier\label{Sec-spike} }

\begin{figure}[ht]
\centering
  \includegraphics[width=3.0in]{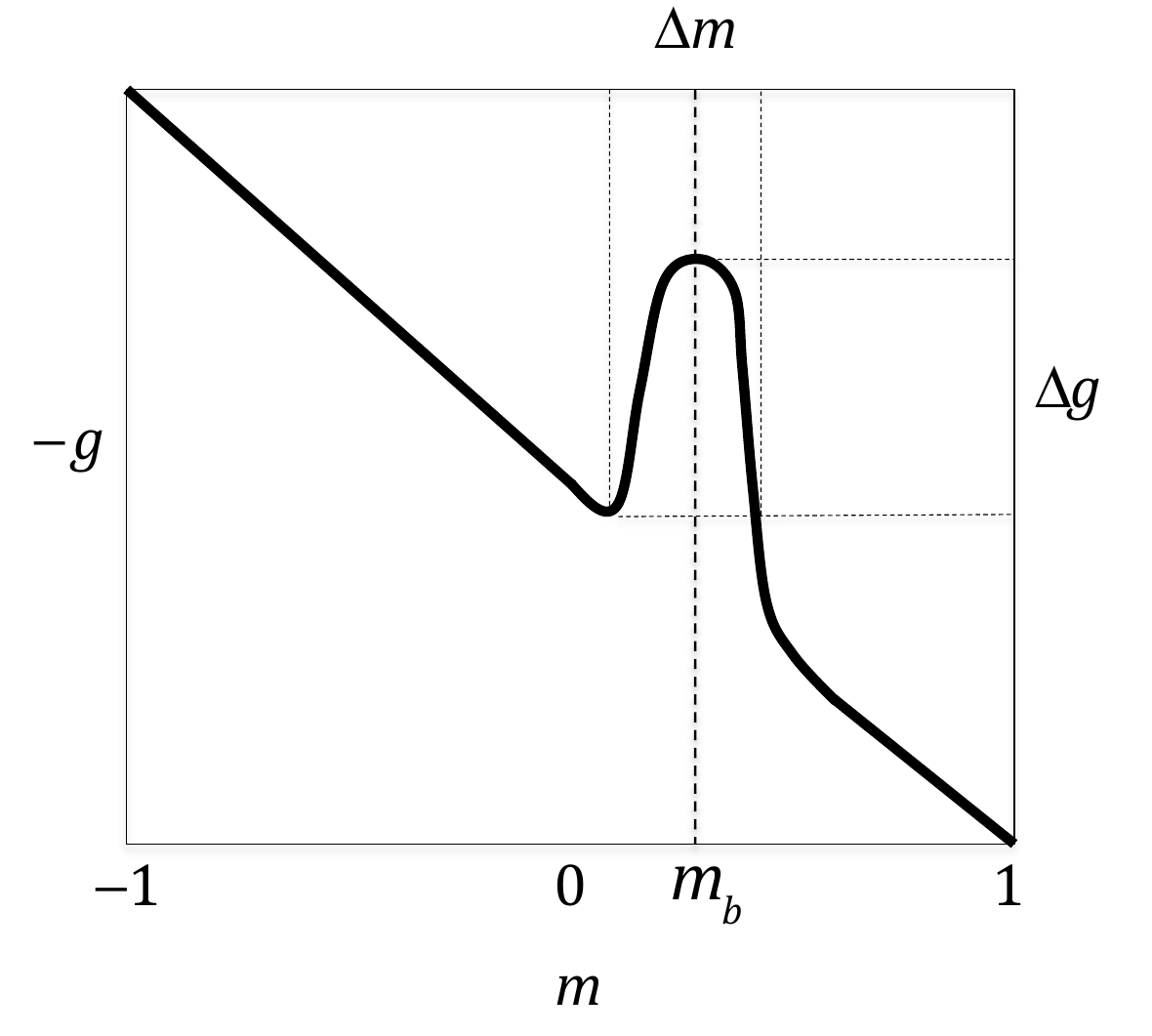}
   \caption{Cartoon of the energy function $-g(m)$ for a problem with small and narrow  barrier.  The barrier dimensions obey  $\Delta m \ll \Delta g \ll 1$ [cf.~(\ref{eq:sc}]. \label{fig:spike}}
\end{figure}

 It is instructive to compare the quantum tunneling rate, or equivalently the classical QMC rate, with the transition rate corresponding to a  classical, purely thermal activation over the barrier.  The latter is the principal mechanism for the transitions between states in simulated annealing (SA) optimization algorithms.  In the present framework, that cost function is  $-g(m)$, which has  multiple minima separated by a barrier.  As an illustrating example, we consider  $-g(m)$ shown in  Fig.~\ref{fig:spike},
 \begin{equation}
-g(m)=-g_0(m)+\Delta g \,f\left(\frac{m-m_b}{\Delta m}\right)\,, \label{eq:spike}
 \end{equation}
 which is monotonically decreasing except for the small region around $m=m_b$,  where there is a narrow spike of the height $\Delta g$ and the typical width $\Delta m$.  Problems of this type were studied previously in \cite{farhi2002quantum,reichardt_quantum_2004,crosson_simulated_2016}.

The  function $f$ describes the shape of the spike (energy barrier), $f^\prime(0)=0, f^{\prime\prime}(0)<0$.  We set the following  relations 
  \begin{equation}
 \Delta g=c N^{-\chi},\quad \Delta m=d N^{-\delta},\quad  \chi < \delta  < 1\label{eq:sc}
 \end{equation}
 where  $c,d={\cal O}(1)$ are constants. The main results  will not depend on the specific form of the functions $g_0$ and $f$  other than  they  are continuous and  
 $|d^kg(m)/dm^k|={\cal O}(1)$ for $k=0,1,2$. We also assume that  $f(q)\rightarrow 0$ exponentially quickly for $|q|\gg 1$.  
 
 The global minimum of the cost function $-g(m)$ corresponds to $m=1$ and there is a local minimum just to the left of the spike at  $m_b-m\ll 1$ as can be seen in Fig.~\ref{fig:spike}. If we initially prepare the system in a state with $m<m_b$, then thermal excitations will cause an over-the-barrier transitions with a rate that scales as $e^{-N \Delta g}$. Therefore, simulated annealing solves the corresponding optimization problem with high probability in time that scales as $\exp(c N^{1-\chi})$.

During QA with the Hamiltonian (\ref{eq:H-sigma}),  the transverse field $\Gamma=\Gamma(t)$ is varied in time. It starts at sufficiently large initial value  of $\Gamma \gg 1$ when all spins are polarized in the $x$ direction. Then, $\Gamma(t)$ is  slowly reduced to zero at the end of the algorithm. We consider the case of  low temperatures where thermal fluctuations can be neglected (see below).
 If the evolution is adiabatic, then the system stays at the instantaneous ground state of $H(t)$ at all times during QA. 
This state corresponds to the irreducible subspace with the maximum total spin $\ell=1$. At the end of QA the system arrives at the global minimum of $-g(m)$ corresponding to all spins pointing in the positive $z$ direction.

The effective potential $U(m,\Gamma)=-\Gamma\sqrt{1-m^2}-g(m)$ [cf. (\ref{eq:Ueff})] is time dependent as $\Gamma(t)$ evolves during QA.  
The extreme points of the potential are obtained from the equation
\begin{equation}
\frac{\partial U}{\partial m}=\frac{m \Gamma}{\sqrt{1-m^2}}-g_{0}^{\prime}(m)+\frac{\Delta g}{\Delta m} f^{\prime}\left(\frac{m-m_b}{\Delta m}\right)=0\;.\label{eq:Uext}
\end{equation}
Following the discussion in Sec.~\ref{sec:WKB} we denote as $m_{0,1}(\Gamma)$ the two instantaneous minima of the effective potential. The maximum of the potential $m_2(\Gamma)$ is very close to  $m_b$ for all $\Gamma$:
\begin{equation}
m_2(\Gamma)-m_b\simeq  \frac{d^2 N^{\chi-2\delta}}{c |f^{\prime\prime}(0)|}\left(  \frac{m_b \Gamma}{1-m_b^2}- g_{0}^{\prime}(m_b)\right )\;.
\end{equation}
 At the beginning of QA  when the term  $\propto \Gamma$ in (\ref{eq:Uext}) is dominating,  the system resides in the vicinity of the global minimum at $m_0=0$.  At the end of QA the system is expected to be  at the global minimum $m_1=1$. Because the barrier is vary narrow, we can find the value $\Gamma=\Gamma_c$ at which the minima exchange orders,
 \begin{equation}
\Gamma_c\simeq g_{0}^{\prime}(m_b)\sqrt{\frac{1}{m_{b}^{2}}-1}\,.\label{eq:Gamma0}
\end{equation}
In the vicinity of this point we have $m_{0,1}(\Gamma_c)\simeq m_b$. The value of the effective potential at the  minima is  $U[m_{0,1}(\Gamma_c)]\simeq -g_0(m_b)$. 

To calculate the tunneling rate the  action under the barrier (\ref{eq:a}) needs to be calculated at  the instanton trajectory with the energy corresponding to  the minima of $U$ at $\Gamma=\Gamma_c$. The maximum value of the momentum $p(m)$ (\ref{eq:kr})  is reached at the 
 \lq\lq middle"  point $m=m_b$ under the barrier and  to  the leading order in $\Delta g$ equals
 \begin{equation}
 p(m_b)=\gamma \Delta g^{1/2},\,\,\,\gamma= \left(g_{0}^{\prime}(m_b)\sinh(\ln m_{b}^{-1})\right)^{1/2}
 \end{equation}
Because $p(m)$ equals zero at the turning points the instantonic action can be estimated as $S=\mu p(m_b) \Delta m$ with $\mu={\cal O}(1)$.  The   factor  $\mu<1$   depends on the shape of the barrier function $f$ (\ref{eq:spike}). For the limiting case of a rectangular barrier $\mu\rightarrow 1$.
The  tunneling transition rate has the form
\begin{equation}
W_{\rm tunn}=B_{\rm tunn}\, e^{-N \gamma \mu   \Delta m \Delta g^{1/2} }\,,
\end{equation}
where  $B_{\rm tunn}$ is a prefactor that scales polynomially with $N$.
 Using (\ref{eq:sc}) we get
\begin{equation}
W_{\rm tunn}=B_{\rm tunn}\, e^{- \kappa   N^{1-\delta-\chi/2}  },\quad \kappa=c^{1/2} d  \gamma \mu\;.\label{eq:Wpsc}
\end{equation}
For the WKB analysis to be valid  the tunneling action (\ref{eq:a}) must obey $S\gg 1$. This implies that one of the conditions must  be true: either $\delta+\frac{\chi}{2} <1$ or $\delta+\frac{\chi}{2}=1$,  $\kappa\gg 1$.

Using the Eqs.~ (\ref{eq:opt}) with $\ell=1$, for  thermal fluctuations to be neglected in QA the temperature must satisfy  the condition
\begin{equation}
T \ll  T_c= \frac{\hbar}{s_0(e_b)},\quad e_b=U(\Gamma_c,m_b) \label{eq:lo}
\end{equation}
where $ 2s_0(e_b)$ [Eq.~(\ref{eq:dade})] is the period of the instanton motion under the barrier  with the energy  close to the barrier top $U(m_b,\Gamma_c)$ and $\Gamma=\Gamma_c$.   It can be immediately seen from the equations for  the instanton velocity (\ref{eq:nuH1}) and (\ref{eq:sc}) that $s_0(e_b)\sim |g^{\prime\prime}(m_b)|^{-1/2}\sim N^{-\delta+\chi/2}$  is decreasing   and therefore  the temperature $T_c$ is increasing with $N$. Therefore,  the rate of the tunneling transition approaches  the zero-temperature limit for any fixed $T$ as $N$ increases.  Because the exponents in the quantum tunneling rate $W_{\rm turn}$ and the QMC transition rate $W_{\rm QMC}$ are identical, the latter is also well approximated by the zero-temperature limit when $T \ll  T_0$. 

Let us assume that the following condition is satisfied
\begin{equation}
\delta+\frac{\chi}{2} =1,\quad \kappa\gg 1,\quad (\chi < \delta <1)\;.\label{eq:win}
\end{equation}
 In this case, both the quantum tunneling and the QMC transition rates can be analyzed using the methods developed in the former sections. They scale only polynomially with $N$, because the exponent in Eq.~(\ref{eq:Wpsc}) does not depend on $N$. However, the rate of the purely thermal transition scales exponentially with $N$. This implies that,
while quantum tunneling is exponentially faster than classical SA algorithms, it still does not offer scaling advantages over classical algorithms for the situations where the path integral is dominated by a single path (instanton). We note that the above conclusion will hold also for a broader range than that given in (\ref{eq:win}):
\begin{equation}
1> \chi \geq  2(1-\delta)\geq 0\;.\label{eq:win1}
\end{equation}
In this case, the QMC transition rate and quantum tunneling rate can only increase compared to the case (\ref{eq:win}) while the thermal transition rate is still exponential.  
An example of this situation is $\chi=0$ and $\delta=1$ corresponding to the barrier in the interaction energy $-N g(m)$ that has for form of Kronecker delta with the height that scales as $N$. However, neither WKB nor Kramers escape theory (for QMC) do work in this case and the actual transition rate expressions  should be analyzed by  different methods.

\section{Discussion and Summary}

In this paper, we considered QMC simulations of {\it thermally assisted} quantum tunneling for an $N$-spin mean-field model. We demonstrated, in a closed analytical form, the equivalence between the exponential scaling of the QMC transition rate and the actual physical tunneling rate with the number of co-tunneling spins.  This equivalence was established numerically  in a previous study~\cite{isakov_understanding_2016} in the effective \lq\lq zero-temperature" limit where the effect of thermal excitations on the tunneling rate can be neglected, and under the condition of zero bias. In this paper, we provided a detailed theoretical description of  the results of \cite{isakov_understanding_2016}  and extended them to  the case of thermally assisted tunneling at finite temperatures and arbitrary biases. We also provided numerical QMC study   to complement theoretical results. 

The findings of the identical scaling of QMC and quantum tunneling seem counterintuitive at a first glance; incoherent tunneling decay is a non-equilibrium process while QMC simulations describe fluctuations around equilibria. In mean-field  spin models, the density of states increases exponentially with energy, therefore providing for a large number of tunneling channels at finite temperature  (\ref{eq:F}).  Both the quantum system and QMC simulation process  thermalize in the metastable domain much faster than a Kramers transition or a tunneling decay occurs. However, it is not clear ahead of time that QMC explores the decay channels in the same way as the quantum system.

The rate of quantum spin tunneling can be written as a product of a polynomial (in $N$) prefactor and an exponential function. The exponential function dominates at the large-$N$ limit, and it can be determined by the change in the effective free-energy functional $\mathfrak{F}[m(\tau)]$ between the values calculated at the instanton and the local minimum [see Eq.~(\ref{eq:Wta})]. Tunneling can thus be described by a most probable path (instanton) corresponding to a single channel that minimizes the effective energy.
  
The stochastic process in the QMC simulation samples the quasi-equilibrium distribution determined by the classical mean-field free-energy functional $\mathfrak{F}[m(\tau)]$ [see Eq.~(\ref{eq:Fm})]. The Kramers escape event in the stochastic process describes the transition from a local minimum to the global minimum, which is dominated by a single \lq\lq transition state"  (a saddle point of $\mathfrak{F}[m(\tau)]$) that the system needs to reach in order to make an escape from the metastable state. 
This transition state corresponds to a quantum instanton, and the change in free energy needed to reach this state is the same as that in the quantum case. This explains the equivalence in the exponential scaling of the QMC transition rate and thermally assisted quantum tunneling rate.
 
We find the solution to the instanton and the change in free energy in a closed analytical form for a general mean-field quantum spin model with the Hamiltonian $ \hat H = -2 \Gamma \hat S_x - N g(2 \hat S_z/N)$. This is achieved by establishing  a  detailed connection between the $\mathfrak{F}[m(\tau)]$ [Eq.~(\ref{eq:Fm})] and the analysis performed with WKB method that deals  explicitly with the system eigenstates and takes into account the degeneracy of the collective spin states with different total spin.

We note that despite a substantial body of work on the models of this  type \cite{boulatov2003quantum,jorg2010energy,seoane2012many,Semerjian-wkb}, the closed analytical  form of the instanton based on the free-energy functional $\m F[m(\tau)]$ has not been obtained previously. 

We think that the spin-instanton method developed in this paper can be generalized to the case of fully connected spin-glass models in transverse field with first-order phase transition such as the $p$-spin models with $p>2$. Specifically, the mathematical approach developed in Sec.~\ref{sec:double} can be used to calculate the instanton solutions in the replicated free energy within the one-step replica symmetry ansatz \cite{jorg2008simple}.


We analyzed the problems with a narrow and tall barrier in the cost function. Assuming that  the  barrier height $N\Delta g ={\cal O}(N^{1-\chi})$ and width  $N \Delta m={\cal O}(N^{1-\delta})$ we established that under the condition  $1> \chi>2(1-\delta)>0$ the quantum tunneling  and QMC rates scale polynomially with $N$ while the rate of purely classical thermal activation over  the barrier scales exponentially with $N$. An interesting case $1-\delta\ll1$ corresponds to a very narrow barrier. In this case, the scaling exponent $1-\chi$  for the barrier height $N\Delta g$  can be very close to 1 for QMC and quantum tunneling rates to be polynomial functions of $N$.

Recently, Brady and van Dam~\cite{brady_quantum_2016} found  numerical evidence that QMC algorithms will succeed in the same regimes where quantum adiabatic optimization succeeds. More recently, Crosson and Harrow~\cite{crosson_simulated_2016} considered a bit-symmetric cost function with a thin, high-energy barrier. They proved that the Markov chain underlying QMC finds the global minimum in polynomial time (in $N$) if the height of the barrier scales less than order $N^{1/2}$. They also conjectured that this is true even for higher barriers that scale as $N$.  

We note that the result obtained in Ref.~~\cite{crosson_simulated_2016} refers to a specific  form  of the interaction energy density $-g(m)$ [Eq.~(\ref{eq:H-sigma})] with a delta-function barrier and represents a particular choice of the broad class of barriers discussed in our paper  where QMC and quantum tunneling  scales polynomially. 

In passing, we  would like to briefly mention several areas open for further investigation where obstructions for the efficient simulation of quantum tunneling with QMC might exist. One of such obstructions is that QMC is not always ergodic. QMC performs dynamics of paths, which is not a representation of the dynamics of the corresponding system. A well-known problem resulting from this is that QMC might have conserved quantities not present in the physical system, such as the number of world lines (particles, magnetization), braiding, or winding numbers~\cite{evertz_loop_2003,hastings}.

QMC may also be less efficient compared to QA in the optimization problems that require multidimensional tunneling to reach the solution. Often in these problems the semiclassical action under the barrier $S({\bf x})$ is not purely imaginary and displays complex features due to the presence of caustics, non-integrability, and non-analyticity. In this case, {\it no} tunneling path can be defined and a Huygens-type wave propagation should be carried out that involves both ${\rm Re}\, S(x)$ and ${\rm Im}\,S(x)$~\cite{huang_wentzel-kramers-brillouin_1990,takada_wentzelkramersbrillouin_1994}. Due to the highly oscillating nature of the wave function $\Psi(x)$ in the classically forbidden region, it is not clear if the associated probability $|\Psi(x)|^2$ can be faithfully recovered with QMC.  
  
It is an important open question as to how QMC will perform in comparison with QA in the problems  that exhibit many-body location and delocalization (MBLD) transitions at finite values of transverse fields \cite{Laumann-EPL-2014}. In the problems with disorder and frustration delocalized states  can exist in the range of energies  with exponential many local minima separated by  large Hamming distances from each other.  A multitude of tunneling paths connects these minima  together and positive interference  gives rise to extended states in the space of spin configurations above the mobility edge \cite{Laumann-Scardicchio-2014}. It is interesting to explore if  the properties of the delocalized phase are important for  the   QA dynamics  towards regions of lower energies where approximate solutions can be obtained in the vicinity of the MBLD transition. In contrast, QMC tunneling, being a classical phenomena, only connects a pair of minima at a time without reproducing the positive interference among exponentially many paths.

An interesting case where QA can have a scaling advantage over classical algorithms refers to the tunneling in non-stoquastic spin Hamiltonians where the negative sign problem prevents a matching QMC algorithm.

Finally, it was found in Ref.~\cite{isakov_understanding_2016} that a version of QMC with open boundary conditions can provide a quadratic speedup compared to incoherent tunneling rate for the Hamiltonian (\ref{eq:H-sigma}) (i.e., the scaling of QMC escape rate with $N$ matches that of quantum tunneling {\it amplitude} instead of the rate). This result can be obtained analytically by a direct extension of the present analysis, which is referred to future studies.

\section{Acknowledgements}

The authors would like to acknowledge support from the NASA Advanced 
Exploration Systems program and NASA Ames Research Center. 
This work was supported in part by 
the  AFRL Information Directorate under grant F4HBKC4162G001,  
the Office of the Director of National Intelligence (ODNI), and 
the Intelligence Advanced Research Projects
Activity (IARPA), via IAA 145483.
The views and conclusions contained herein
are those of the authors and should not be interpreted as necessarily
representing the official policies or endorsements, either expressed
or implied, of ODNI, IARPA, AFRL, or the U.S. Government.  The U.S.
Government is authorized to reproduce and distribute reprints for
Governmental purpose notwithstanding any copyright annotation thereon.


%

\appendix

\vspace{0.4in}
\begin{center}
{\Large\bf Appendix}
\end{center}
\hfill

In this appendix, we show that the  static free energy per spin $\mathfrak{F}_0$ that appears in (\ref{eq:Wta}) equals to the free energy density $F(m_0)$ from (\ref{eq:WQMC}). 
For the static solution $m(\tau)=m_0$ corresponding to the local minimum of the QMC free energy (\ref{eq:static_free_energy}), the propagator Eq.~(\ref{eq:propagator}) becomes
\begin{align}
 K^{\beta, 0} = e^{\beta [\Gamma \sigma_x+\g'(m_0) \sigma_z]}\;.
\end{align}
Consequently, the self-consistent condition~(\ref{eq:l_self}) becomes
\begin{align}\label{eq:l_qmc}
\begin{split}
 \scl  &= \tanh \Big(\beta \sqrt {\Gamma^2 + \big(\g'(m_0)\big)^2}\;\Big)\;.
 \end{split}
\end{align}
The condition for extreme points of the static free energy Eq.~(\ref{eq:static_free_energy}) is
\begin{align}\label{eq:extreme_point_static}
 m \sqrt{\Gamma^2 + \big(\g'(m)\big)^2} - \g'(m)\ell = 0\;,
\end{align}
where we use Eq.~(\ref{eq:l_qmc}) to simplify the above expression. 
Solving $\g'(m)$ from the above equation, we have
\begin{align}\label{eq:qmc_condition_b}
 \g'(m) = \Gamma m\, (\ell^2-m^2)^{-1/2}\;.
\end{align}
Thus, the pair of equations~(\ref{eq:l_qmc}) and (\ref{eq:qmc_condition_b}) determine the local minimum $m=m_0$ and other extreme points of the \lq\lq static" free energy (\ref{eq:static_free_energy}). 

In the WKB approach, the effective potential $U_\ell(m)=\varepsilon_\ell(m,0)$ (\ref{eq:Hcl}) takes the form
\begin{align}
U_\ell(m) = -\Gamma \sqrt{\ell^2-m^2} -\g(m)\:. 
\end{align}
The extreme points of the effective potential satisfy
\begin{align}
\g'(m) = \Gamma m\, (\ell^2-m^2)^{-1/2}\;,\label{eq:wkb_condition_b}
\end{align}
which is the same as Eq.~(\ref{eq:qmc_condition_b}). We will use the following equivalent form of Eq.~(\ref{eq:wkb_condition_b}),
\begin{align}\label{eq:wkb_condition_c}
\begin{split}
 \Gamma^2 + \big(\g'(m)\big)^2  &= \Gamma^2 l^2\, (\ell^2-m^2)^{-1}\;.
\end{split}
\end{align}
According to Eq.~(\ref{eq:l0}) the optimal values of $\ell$ satisfy
\begin{align}\label{eq:wkb_condition_a}
\frac{\dif Q_\ell}{\dif \ell} = \beta\, \frac{\partial U_\ell(m)} {\partial \ell} = -\beta \, \Gamma \ell\, (\ell^2-m^2)^{-1/2}\;,
\end{align}
where $Q_\ell$ is the binary entropy. Putting Eq.~(\ref{eq:wkb_condition_c}) into Eq.~(\ref{eq:wkb_condition_a}), we have
\begin{align}
\begin{split}\label{eq:extreme_l_stationary}
 \frac{\dif Q_\ell}{\dif \ell}  &= -\beta \sqrt{\Gamma^2 + \big(\g'(m)\big)^2}\;,
 \end{split}
\end{align}
which is equivalent to Eq.~(\ref{eq:l_qmc}) by using the definition of the entropic factor $Q_\ell$ in  Eq.~(\ref{eq:Q}).  Thus, we have shown that the extreme points of the stationary solutions for QMC and WKB are the same.

Using the conditions (\ref{eq:l_qmc}) and (\ref{eq:qmc_condition_b}), the QMC free energy (\ref{eq:static_free_energy}) at the extremum takes the form 
\begin{align}
 \frerg_0 &= \Gamma m^2\, (\ell^2-m^2)^{-1/2} -\g(m)  - \frac{1}{\beta} \ln \frac{2}{\sqrt{1-\ell^2}}\;.\label{eq:free_erg_static_y}
\end{align}
From the definition of the entropy Eq.~(\ref{eq:Q}), we have
\begin{align}\label{eq:entropy_identity}
 \ln \frac{2}{\sqrt{1-\ell^2}} = Q_\ell + \ell \tan^{-1}\ell \;.
\end{align}
From Eqs.~(\ref{eq:l_qmc}), (\ref{eq:extreme_point_static}), and (\ref{eq:qmc_condition_b}), we also have
\begin{align}\label{eq:another_identity}
\tan^{-1}\ell= \beta \Gamma \ell\, (\ell^2-m^2)^{-1/2} \;.
\end{align}
Putting Eq.~(\ref{eq:another_identity}) into Eq.~(\ref{eq:entropy_identity}) and then putting the result into Eq.~(\ref{eq:free_erg_static_y}), we have
\begin{align}
 \frerg_0 =F(m_0) &= -\Gamma \sqrt{\ell^2-m^2}-\g(m)-\frac{1}{\beta}\,Q(\ell)\;,\label{eq:free_erg_static_z}
\end{align}
which is exactly the static WKB free energy $\beta \mathfrak{F}_0$ given that we are using the optimal values of $m=m_0$ and $\ell$ as described above.

\end{document}